\pdfoutput=1
\documentclass[aps,pre,preprint,superscriptaddress,amsmath,showkeys]{revtex4-1}
\usepackage[hyperindex,breaklinks,colorlinks=true]{hyperref}
\usepackage{graphicx}
\usepackage[T1]{fontenc}
\usepackage[utf8]{inputenc}

\newcommand{\llangle}{\left\langle}
\newcommand{\rrangle}{\right\rangle}
\newcommand{\hrho}{\hat\rho}
\newcommand{\brho}{\boldsymbol{\rho}}
\newcommand{\hbrho}{\hat{\boldsymbol{\rho}}}
\newcommand{\D}{\mathcal{D}}
\newcommand{\rmd}{\mathrm{d}}

\begin{document}
\title{Extreme current fluctuations of boundary-driven systems in the large-$N$ limit}

\author{Yongjoo Baek}
\email{yongjoo.baek@physics.technion.ac.il}
\affiliation{Department of Physics, Technion, Haifa 32000, Israel}

\author{Yariv Kafri}
\affiliation{Department of Physics, Technion, Haifa 32000, Israel}

\author{Vivien Lecomte}
\affiliation{Laboratoire Probabilit\'{e}s et Mod\`{e}les Al\'{e}atoires, UMR7599 CNRS, Sorbonne Paris Cit\'e, Universit\'{e} Pierre et Marie Curie \& Universit\'{e} Paris Diderot, F-75013 Paris, France}

\date{\today}

\begin{abstract}
Current fluctuations in boundary-driven diffusive systems are, in many cases, studied using hydrodynamic theories. Their predictions are then expected to be valid for currents which scale inversely with the system size. To study this question in detail, we introduce a class of large-$N$ models of one-dimensional boundary-driven diffusive systems, whose current large deviation functions are exactly derivable for any finite number of sites. Surprisingly, we find that for some systems the predictions of the hydrodynamic theory may hold well beyond their naive regime of validity. Specifically, we show that, while a symmetric partial exclusion process exhibits non-hydrodynamic behaviors sufficiently far beyond the naive hydrodynamic regime, a symmetric inclusion process is well described by the hydrodynamic theory for arbitrarily large currents. We conjecture, and verify for zero-range processes, that the hydrodynamic theory captures the statistics of arbitrarily large currents for all models where the mobility coefficient as a function of density is unbounded from above. 
    In addition, for the large-$N$ models, we prove the additivity principle under the assumption that the large deviation function has no discontinuous transitions.
\end{abstract}


\maketitle

\tableofcontents

\section{Introduction} \label{sec:intro}

One of the most fundamental ways to characterize the steady state of a system is through the statistical properties of currents. These have been studied both in and out of equilibrium and in both classical~\cite{Derrida2007} and quantum systems~\cite{pilgram_stochastic_2003,Esposito2009,genway_trajectory_2014}. Recently, much progress has been achieved in understanding the statistics of time-averaged currents, which are encoded in a corresponding large deviation functions (LDF), of boundary-driven diffusive systems in one dimension~\cite{Derrida2004,Bodineau2004,Bertini2005a,Bertini2006,Harris2005,Imparato2009,Lecomte2010,Shpielberg2015} as well as in other geometries~\cite{bodineau_vortices_2008,Akkermans2013}. Whereas exact microscopic solutions are often available for bulk-driven systems~\cite{derrida_exact_1998,prolhac_cumulants_2009,lazarescu_exact_2011,mallick_exact_2011,lazarescu_matrix_2013,lazarescu_physicists_2015,ayyer_full_2015},
the results for boundary-driven systems largely rest on the application of a hydrodynamic approach termed the macroscopic fluctuation theory (MFT)~\cite{spohn_large_1991,Bertini2002,jordan_fluctuation_2004,Bertini2015}, with the notable exception of~\cite{lazarescu_matrix_2013}.

Being a hydrodynamic theory, the MFT is naively expected to yield the correct statistics of currents only when the current fluctuations are small enough for the hydrodynamic description to be valid. For example, consider a single-species diffusive system on the line $0 \leq x \leq \ell$, where $\ell$ denotes the length of the system. After coarse-graining and a diffusive rescaling ($x \to x/\ell$ and $t \to t/\ell^2$~\footnote{These notations indicate that the rescaled variables are defined as $\tilde{x} \equiv x/\ell$ and $\tilde{t} \equiv t/\ell^2$, and then renamed as $x$ and $t$, respectively. Other notations for rescaling schemes should be interpreted similarly.}), the hydrodynamic equation takes the form
\begin{equation}
	\label{eq:hydro}
	\partial_t \rho(x) = -\partial_x J(x),
\end{equation}
with $\rho(x)$ the coarse-grained density and $J(x)$ the coarse-grained current. Since we are interested in the $\ell \to \infty$ limit, this equation is not well defined for $J(x)$ which before the rescaling is not of the order of $1/\ell$. Thus, the statistics of currents obtained by the MFT are reliable only for current fluctuations of the order of $1/\ell$. The same conclusion can be reached by another argument more directly based on the MFT, which is discussed in Appendix~\ref{app:hydro_limit}.

In this paper we study the validity of the hydrodynamic approach in regions where it is expected to fail. Quite surprisingly, we find that there are classes of models where the hydrodynamic approach captures the statistics of currents much beyond its naive regime of validity. We give a simple explanation for this phenomena and based on it argue that this behavior is expected to be generic when the mobility diverges with the density of particles.

To obtain these results, we study current LDFs of boundary-driven systems whose lattice structure is preserved, keeping a finite number of sites $L$. Since the exact current LDFs of microscopic lattice models are difficult to obtain (with the exception of the zero-range-process~\cite{Harris2005}), we consider a little-studied class of coarse-grained models, which we term {\em large-$N$ models}. A large-$N$ model consists of a one-dimensional chain of boxes, each of which holds a macroscopically large number of particles (controlled by $N$) and which relaxes instantaneously to local equilibrium. As such, it retains the lattice structure even after coarse-graining and can be thought of as an analog of the ``boxed models'' studied in~\cite{Bunin2013,Kafri2015}. In a manner similar to models of population dynamics~\cite{Elgart2004,Meerson2011} and lattice spin models in the large-spin limit~\cite{Tailleur2007,Tailleur2008}, we rescale dynamical variables and hopping rates of the model by powers of $N$. This allows us to apply the standard saddle-point techniques in the $N \to \infty$ limit.

Thanks to simplifications arising from the assumption of a macroscopic number of particles at each box (site), the current LDFs of our large-$N$ models are exactly derivable even for a finite system with any number of sites $L$. By comparing the tail behaviors of the current LDFs in the large-$L$ limit with the predictions of the MFT approach, we can observe how and when non-hydrodynamic behaviors start to emerge. Interestingly, our formulation also shows that the same microscopic dynamics may produce {\em different macroscopic models} depending on how the microscopic variables are scaled with $N$.

We note that there were previous studies on models with multiple particles per site, such as partial exclusion processes~\cite{schutz_non-abelian_1994}, inclusion processes~\cite{giardina_duality_2007}, or both~\cite{giardina_correlation_2010,carinci_duality_2013}. These studies obtained exact expressions for particle density correlations on a finite lattice with $L$ sites. The corresponding density large deviations were studied in~\cite{Tailleur2007,Tailleur2008}, but only after a gradient expansion in the $L \to \infty$ limit that washes away the lattice structure. To our knowledge, large deviation properties of these models at finite $L$ have not been properly explored~\footnote{We note that there was a previous attempt to calculate the current LDF of a discrete system by applying a saddle-point approximation directly to the microscopic model~\cite{Imparato2009}. This approximation, however, is not well controlled.}.

This paper is organized as follows. In Sec.~\ref{sec:models}, we introduce two classes of large-$N$ models, which are the symmetric partial exclusion process (SPEP) and the symmetric inclusion process (SIP). It is shown that the latter becomes equivalent to the well-studied Kipnis--Marchioro--Presutti (KMP) model~\cite{Kipnis1982,Bertini2005b} after an appropriate rescaling by $N$. In Sec.~\ref{sec:spep_current}, we study current large deviations of the SPEP, which exhibits non-hydrodynamic behaviors for current fluctuations sufficiently far beyond the naive hydrodynamic regime expected by the argument given above. In addition, we also discuss the validity of the additivity principle. In Sec.~\ref{sec:sip_current}, we analyze current large deviations of the SIP for different large-$N$ limits, which in all cases exhibit hydrodynamic behaviors for arbitrarily large current fluctuations. Based on these results, in Sec.~\ref{sec:criterion} we propose a criterion for the persistence of hydrodynamic current fluctuations in the non-hydrodynamic regime, and confirm its validity for the symmetric zero-range process. Finally, we summarize our results and conclude in Sec.~\ref{sec:conclusions}.

\section{Large-$N$ models} \label{sec:models}

We now turn to introduce the large-$N$ versions of the SPEP and the SIP. Starting with the SPEP the microscopic model is defined and used to obtain a path-integral representation for the current cumulant generating function (CGF) along with the prescription for calculating it in the large-$N$ limit. The hydrodynamic limit of the model is then presented for completeness. The section closes by giving the corresponding results for the class of SIP models.

\subsection{Microscopic dynamics} \label{ssec:models_micro}

\begin{figure}[b]
	\includegraphics[width = 0.49\textwidth]{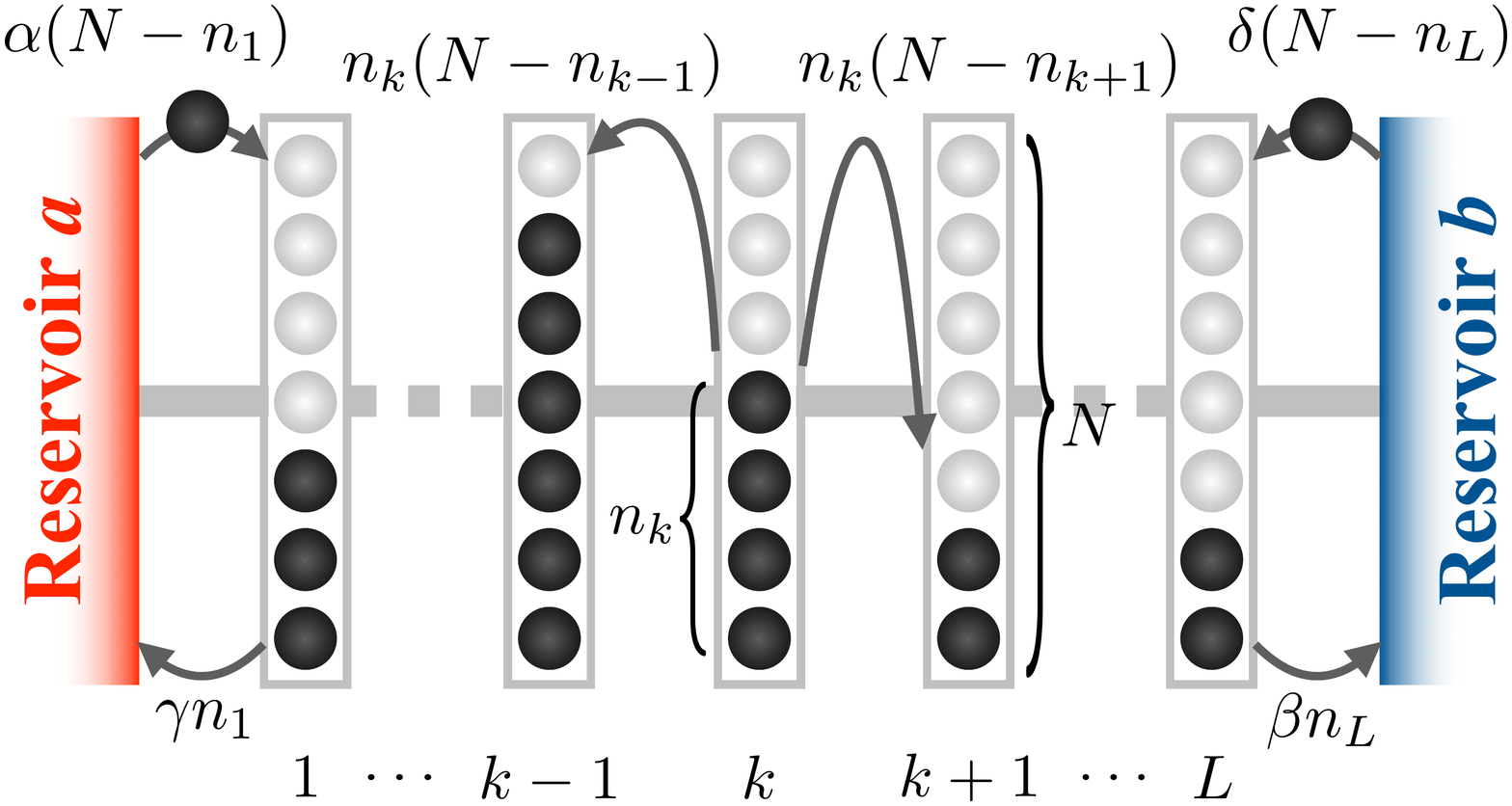}
	\includegraphics[width = 0.49\textwidth]{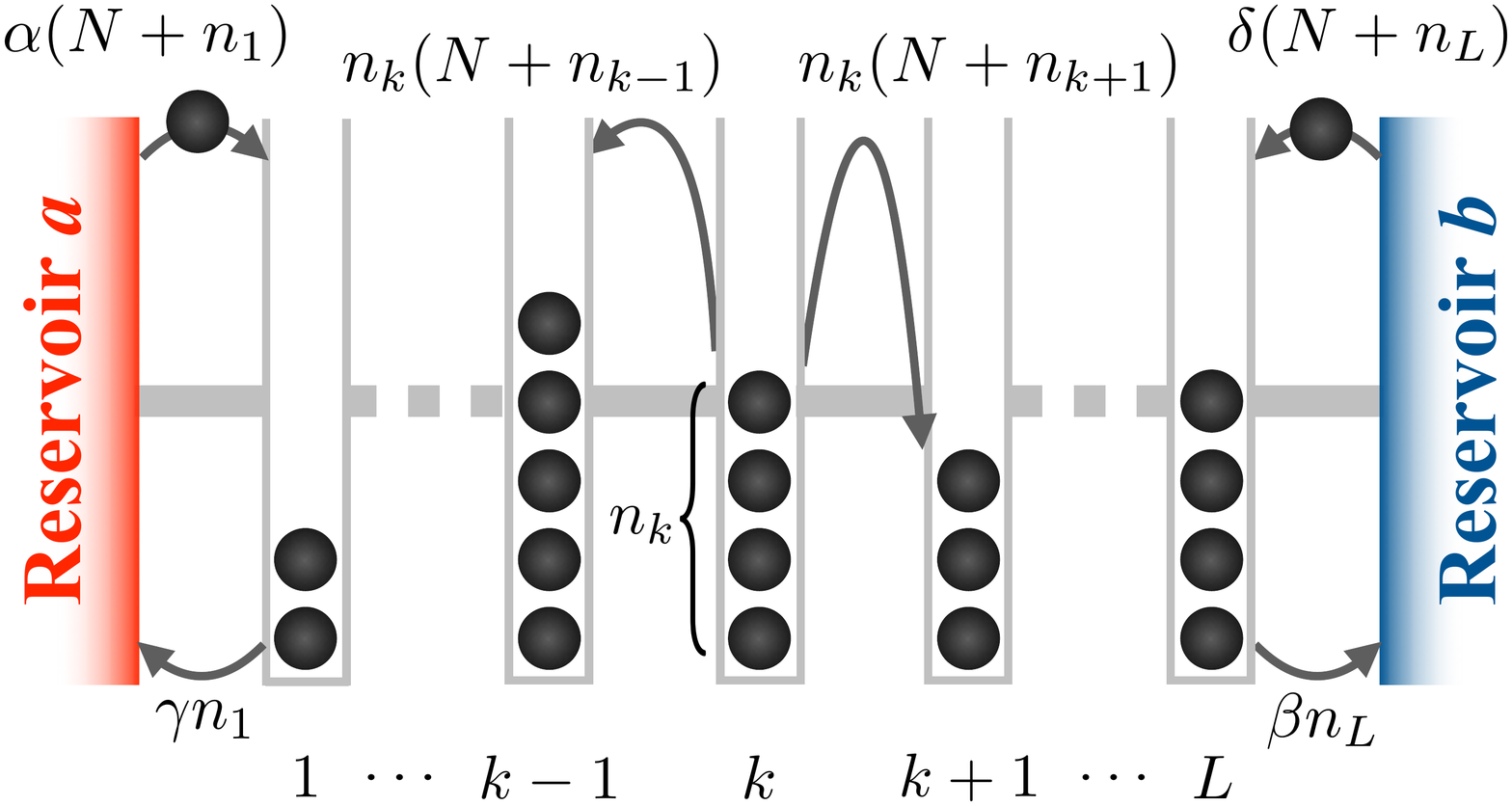}
	\caption{\label{fig:models} Illustrations of two types of large-$N$ models. (Left) The SPEP features repulsive interactions, and each box can hold at most $N$ particles. (Right) The SIP features attractive interactions, and there is no upper bound on the number of particles in each box.}
\end{figure}

The models are defined on a one-dimensional chain of $L$ boxes which are in contact with two particle reservoirs denoted by $a$ and $b$ (see Fig.~\ref{fig:models} for an illustration). Each box is assumed to be in local equilibrium so that the state of box $k$ is completely specified by the number of particles $n_k$, for $k = 1,\,2,\,\ldots,\,L$. A particle hops from a box to an adjacent one with a rate (in arbitrary units) given by
\begin{align}\label{eq:micro_bulk_rates}
	\text{SPEP:} &\qquad (n_k,\,n_l) \xrightarrow{n_k(N - n_l)} (n_k - 1,\, n_l + 1) \qquad \text{for $l = k \pm 1$}, \nonumber\\
	\text{SIP:} &\qquad (n_k,\,n_l) \xrightarrow{n_k(N + n_l)} (n_k - 1,\, n_l + 1) \qquad \text{for $l = k \pm 1$},
\end{align}
which reflects exclusion (`attractive') interactions between particles in the SPEP (SIP). It is clear that for the SPEP the range of $n_k$ is bounded from above and below ($0 \le n_k \le N$), while for the SIP $n_k$ is only bounded from below ($n_k \ge 0$). The hopping rates at the boundaries are defined similarly as:
\begin{alignat}{3}\label{eq:micro_boundary_rates}
	\text{SPEP:} &\qquad n_1 &\xrightarrow{\alpha (N - n_1)} n_1 + 1, &\qquad n_1 &\xrightarrow{\gamma n_1} n_1 - 1, \nonumber\\
	&\qquad n_L &\xrightarrow{\delta (N - n_L)} n_1 + 1, &\qquad n_L &\xrightarrow{\beta n_L} n_L - 1, \nonumber\\
	\text{SIP:} &\qquad n_1 &\xrightarrow{\alpha (N + n_1)} n_1 + 1, &\qquad n_1 &\xrightarrow{\gamma n_1} n_1 - 1, \nonumber\\
	&\qquad n_L &\xrightarrow{\delta (N + n_L)} n_1 + 1, &\qquad n_L &\xrightarrow{\beta n_L} n_L - 1.
\end{alignat}
If the system is coupled only to reservoir $a$ (reservoir $b$), the average number of particles in each box relaxes to $\bar n_a$ ($\bar n_b$) as determined by $\alpha$ and $\gamma$ ($\beta$ and $\delta$). 
In what follows, we fix the contact rates to the reservoirs through $N/(\gamma + \alpha) = 1$, $N/(\beta + \delta) = 1$ for the SPEP, and $N/(\gamma - \alpha) = 1$, $N/(\beta - \delta) = 1$ for the SIP.
The parameters $\bar{n}_a$ and $\bar{n}_b$ thus fully describe the coupling with the reservoirs:
\begin{align} \label{eq:micro_boundary_densities}
	\text{SPEP:} &\qquad \alpha = \bar{n}_a, \quad \beta = N - \bar{n}_b, \quad \gamma = N - \bar{n}_a, \quad \delta = \bar{n}_b \;, \nonumber\\
	\text{SIP:} &\qquad \alpha = \bar{n}_a, \quad \beta = N + \bar{n}_b, \quad \gamma = N + \bar{n}_a, \quad \delta = \bar{n}_b \;.
\end{align}
This choice provides simpler expressions in the results presented below, without affecting the large-$L$ hydrodynamic behavior.

With these definitions it is natural to introduce density variables according to
\begin{equation} \label{eq:density_rescaling}
	\rho_k \equiv \frac{n_k}{N}, \quad \bar\rho_a \equiv \frac{\bar n_a}{N}, \quad \bar\rho_b \equiv \frac{\bar n_b}{N}\;,
\end{equation}
and rescale time as $t \to Nt$. Then the evolution of the average density profile, taken over some initial distribution and denoted by angular brackets, satisfies
\begin{equation} \label{eq:average_dyn}
	\frac{\partial \langle \rho_k \rangle}{\partial t} = \langle \rho_{k-1}\rangle - 2 \langle \rho_k \rangle + \langle \rho_{k+1} \rangle
\end{equation}
for any $k  = 1,\,2,\,\ldots,\,L$ with $\rho_0 \equiv \bar\rho_a$ and $\rho_{L+1} \equiv \bar\rho_b$.
We note that the discrete diffusion equation~\eqref{eq:average_dyn} is also known to hold exactly for the standard Symmetric Simple Exclusion Process (SSEP), which corresponds to the SPEP with $N = 1$.

Under this rescaling, for the SPEP, $N$ is naturally interpreted as the capacity of each box. On the other hand, for the SIP the number of particles is not bounded from above. Therefore, $N$ does not admit a natural interpretation without specifying how both ${\bar n_a}$ and ${\bar n_b}$ scale with $N$. In fact, one can choose an alternate scaling and define densities for the SIP as 
\begin{equation} \label{eq:density_rescaling_alpha}
	\rho_k \equiv \frac{n_k}{N^{1+\alpha}}, \quad \bar\rho_a \equiv \frac{\bar n_a}{N^{1+\alpha}}, \quad \bar\rho_b \equiv \frac{\bar n_b}{N^{1+\alpha}}
\end{equation}
with $t$ rescaled by $N$ as above and $\alpha>0$ (the rationale behind this constraint will become clear below). It is straightforward to check that \eqref{eq:average_dyn} is then unchanged. Interestingly, these two scaling choices for the SIP, as we show below, lead to different {\em macroscopic} theories. In what follows, when we also study the SIP rescaled by \eqref{eq:density_rescaling_alpha} and refer to it as SIP(1+$\alpha$), in contrast to the SIP(1) whose scaling is defined in \eqref{eq:density_rescaling}.


\subsection{SPEP -- current CGF and hydrodynamic limit}\label{ssec:models_spep}

Our interest is in calculating the current CGF which encodes the statistics of the time-averaged density current $J$. We can obtain $J$, for example, by measuring the flux of particles from box $L$ to reservoir $b$ during an interval $t \in [0,\,T]$. The CGF is then defined through
\begin{equation} \label{eq:cgf_def}
	e^{NT\psi_{N,L}(\lambda,\bar\rho_a,\bar\rho_b)} = \llangle e^{N\lambda T J}\rrangle \quad \text{for $T \gg 1$,}
\end{equation}
where the average, denoted by angular brackets, is taken with fixed $\bar\rho_a$ and $\bar\rho_b$, and $\lambda$ is conjugate to the current $J$. Using standard methods (see Appendix~\ref{app:path}), we can write a path-integral representation of the CGF 
\begin{equation} \label{eq:macro_path_integ}
	e^{NT\psi_{N,L}(\lambda)} = \int \D\brho \D\hbrho \,
		\exp\left\{-N\int_{0}^{T} \mathrm{d}t \,\left[ \hbrho\cdot\dot{\brho}
			- H_L(\lambda;\brho,\hbrho)\right] + o(N)\right\}
\end{equation}
with $\brho \equiv (\rho_1, \rho_2, \ldots, \rho_L)$ the density vector and $\hbrho\equiv (\hat\rho_1, \hat\rho_2, \ldots, \hat\rho_L)$ the auxiliary `momentum' vector. For the SPEP, the Hamiltonian $H_L$ is given by
\begin{align}\label{eq:spep_H}
	H^\mathrm{SPEP}_L(\lambda;\brho,\hbrho)
		&= \sum_{k=1}^{L-1} \left[ \rho_k (1-\rho_{k+1}) \left(e^{\hat{\rho}_{k+1} - \hat{\rho}_k}-1\right)
			+ \rho_{k+1} (1-\rho_k) \left(e^{\hat{\rho}_k - \hat{\rho}_{k+1}}-1\right)\right] \nonumber\\
		&\quad +\rho_1 (1-\bar{\rho}_a) \left(e^{- \hat{\rho}_1}-1\right) + \bar{\rho}_a (1-\rho_1) \left(e^{\hat{\rho}_1}-1\right) \nonumber\\
		&\quad +\rho_L (1-\bar{\rho}_b) \left(e^{- \hat{\rho}_L + \lambda}-1\right) + \bar{\rho}_b (1-\rho_L) \left(e^{\hat{\rho}_L - \lambda}-1\right).
\end{align}
When $N$ is very large (in the sense of $N \gg T \gg 1$), the large-$N$ CGF $\psi_L$ can be obtained using saddle-point asymptotics
\begin{equation} \label{eq:cgf_saddle}
	\psi_L(\lambda) \equiv \lim_{N\to\infty} \psi_{N,L}(\lambda) = \lim_{T \to \infty} \frac{1}{T} \inf_{\brho,\,\hbrho}\int_{0}^{T} \mathrm{d}t \,
		\left[ \hbrho\cdot\dot{\brho} - H_L(\lambda;\brho,\hbrho) \right]
\end{equation}
with the infimum taken over trajectories of $\brho$ and $\hbrho$. As advertised above, this approximation requires only $N$ to be a large parameter, so its predictions hold for any value of $L$.
The minimization principle~\eqref{eq:cgf_saddle} is similar to that of the MFT approach~\cite{Bertini2015} for the SSEP, with $N$, instead of $L$, playing the role of the large parameter governing the saddle-point. This allows us to keep track of the lattice structure at any finite~$L$.

Assuming that the minimizing trajectory is time-independent, the saddle-point equations are given by
\begin{equation} \label{eq:time_indep_sol}
	\frac{\partial \brho}{\partial t} = \frac{\partial H_L}{\partial \hbrho} = 0, \quad \frac{\partial \hbrho}{\partial t} = -\frac{\partial H_L}{\partial \brho} = 0.
\end{equation}
The solutions of these equations, which we denote by $\brho^*$ and $\hbrho^*$, are typically called the {\em optimal profiles} which support the current fluctuation $J$. Then the current CGF is obtained from \eqref{eq:cgf_saddle} as
\begin{equation} \label{eq:cgf_H}
	\psi_L(\lambda) = H_L(\lambda;\brho^*,\hbrho^*).
\end{equation}
The additivity principle, proposed in~\cite{Bodineau2004} (also independently studied in \cite{jordan_fluctuation_2004}), implies that the above assumption is applicable for any value of $\lambda$. Although counterexamples were found in periodic bulk-driven systems~\cite{Bertini2005a,bodineau_distribution_2005,Bertini2006,Hurtado2011,Espigares2013}, the principle was analytically shown to be true for any open boundary-driven diffusive system with a constant diffusion coefficient and a quadratic mobility coefficient~\cite{Imparato2009} ---~without ruling out possible discontinuous transitions, which in turn were numerically discarded in~\cite{Hurtado2009} for a specific model related to the SIP. As shown below, both the SPEP and the SIP correspond to this class of systems in the hydrodynamic limit. Thus we expect that the same principle is also applicable to our large-$N$ models, and discuss arguments supporting its validity in Sec.~\ref{ssec:spep_fin_N} and Appendix~\ref{app:finiteNquantumfluctuationsSPEP}.

Finally, we show that under appropriate assumptions our large-$N$ models are well described by hydrodynamic theories. To see this, we first apply a diffusive scaling in terms of $L$, which involves writing the position of box $k$ as $x \equiv k/(L+1)$ (with the lattice spacing set to one) and rescaling time by $t \to t/(L+1)^2$. We also assume that differences between adjacent boxes, namely $\rho_{k+1} - \rho_k$ and $\hrho_{k+1} - \hrho_k$, scale as $1/(L+1)$. Then, in the $L \to \infty$ limit, the gradients $\partial_x \rho$ and $\partial_x \hrho$ are well defined, and \eqref{eq:macro_path_integ} can be approximated as
\begin{equation} \label{eq:hydro_path_integ}
	e^{N (L+1)^2 T\psi(\lambda)} = \int \D \rho \D\hat\rho \,
		\exp\left\{-N(L+1) \int_{0}^{T} \mathrm{d}t \left[ \left( \int_0^1 \mathrm{d}x \, \hrho \dot{\rho} \right) - H [\rho,\hrho] \right]\right\}.
\end{equation}
Here the Hamiltonian $H [\rho,\hrho]$, which is no longer dependent on $\lambda$, is now a functional of continuous profiles $\rho(x)$ and $\hrho(x)$. The functional typically has the form of
\begin{equation} \label{eq:H_mft}
	H[\rho,\hrho] = \int_0^1 \mathrm{d}x \,
			\left[ -D(\rho)(\partial_x \rho)(\partial_x \hrho) + \frac{\sigma(\rho) (\partial_x \hrho)^2}{2} \right]
\end{equation}
with $D(\rho)$ the diffusion coefficient and $\sigma(\rho)$ the mobility coefficient. For the SPEP, these coefficients are given by
\begin{equation}
	D(\rho) = 1, \quad \sigma(\rho) = 2\rho(1-\rho),
\end{equation}
respectively. We note that this $\sigma(\rho)$ is bounded from above, with the maximum value given by $\sigma(1/2) = 1/2$. Meanwhile, the rescaling of time speeds up the microscopic dynamics, so the leftmost ($k = 1$) and rightmost ($k = L$) boxes equilibrate with the coupled reservoirs (see \emph{e.g.}~Appendix B.2 of Ref.~\cite{Tailleur2008}). Hence, the spatial boundary conditions are given by
\begin{equation}
	\rho(0) = \bar{\rho}_a, \quad \rho(1) = \bar{\rho}_b, \quad \hrho(0) = 0, \quad \hrho(1) = \lambda,
\end{equation}
whose dependence on $\lambda$ keeps $\psi$ a function of $\lambda$.

In what follows we list the corresponding sets of results for the SIP(1) and the SIP(1+$\alpha$).

\subsection{SIP(1) -- current CGF and hydrodynamic limit}\label{ssec:models_sip}

It is straightforward to repeat the above derivations for the SIP. We find, using the notation of \eqref{eq:macro_path_integ}, 
\begin{align} \label{eq:sip1_H}
	H^{\mathrm{SIP}(1)}_L (\lambda;\brho,\hbrho) &\equiv \sum_{k=1}^{L-1} \left[ \rho_k (1 + \rho_{k+1})
			\left(e^{\hrho_{k+1} - \hrho_k}-1\right) + \rho_{k+1} (1 + \rho_k) \left(e^{\hrho_k - \hrho_{k+1}}-1\right)\right] \nonumber\\
			&\quad+ \left[ \rho_1 (1 + \bar{\rho}_a) \left(e^{- \hrho_1}-1\right) + \bar{\rho}_a (1 + \rho_1) \left(e^{\hrho_1}-1\right)\right] \nonumber\\
			&\quad+ \left[ \rho_L (1 + \bar{\rho}_b) \left(e^{- \hrho_L+\lambda}-1\right) + \bar{\rho}_b (1 + \rho_L) \left(e^{\hrho_L-\lambda}-1\right)\right].
\end{align}
In addition, the corresponding hydrodynamic Hamiltonian in the large-$L$ limit is given by \eqref{eq:H_mft} with
\begin{equation} \label{eq:coeff_sip1}
	D(\rho) = 1, \quad \sigma(\rho) = 2\rho(1+\rho).
\end{equation}
We note that $\sigma(\rho)$ in this case is not bounded from above.

\subsection{SIP(1+$\alpha$) -- current CGF and hydrodynamic limit}\label{ssec:models_sip1a}

For the SIP(1+$\alpha$) we similarly find, using the notation of \eqref{eq:macro_path_integ}, 
\begin{align} \label{eq:sip1a_H}
	H^{\mathrm{SIP}(1+\alpha)}_L(\lambda;\brho,\hbrho)
		&\equiv \sum_{k=1}^{L-1} \left[-(\rho_{k+1} - \rho_k)(\hrho_{k+1} - \hrho_k) + \rho_k \rho_{k+1}(\hrho_{k+1} - \hrho_k)^2\right] \nonumber\\
		&\quad + (\bar\rho_a - \rho_1)\hrho_1 + \bar\rho_a \rho_1 \hrho_1^2 + (\bar\rho_b - \rho_L)(\hrho_L-\lambda) + \rho_L \bar\rho_b (\hrho_L-\lambda)^2.
\end{align}
The hydrodynamic description of this model in the large-$L$ limit is given by \eqref{eq:H_mft} with
\begin{equation} \label{eq:coeff_sip1a}
	D(\rho) = 1, \quad \sigma(\rho) = 2\rho^2,
\end{equation}
where $\sigma(\rho)$ is again not bounded from above. These transport coefficients are also shared by the Kipnis--Marchioro--Presutti (KMP) model of heat conduction~\cite{Kipnis1982,Bertini2005b}. It is notable that the same microscopic model produces different macroscopic behaviors depending on the reservoir properties.

\section{Current large deviations in the SPEP}
\label{sec:spep_current}

In what follows we first show that the scaled CGF of the time-averaged current in the SPEP in the large-$N$ limit is given by
\begin{equation} \label{eq:spep_cgf}
	\psi^\mathrm{SPEP}_L(\lambda) = \begin{cases}
		(L+1) \sinh^2 \left(\frac{1}{L+1} \mathrm{arcsinh} \sqrt{\omega^\mathrm{SPEP}}\right) &\text{if $\omega^\mathrm{SPEP} \ge 0$}, \\
		-(L+1) \sin^2 \left(\frac{1}{L+1} \mathrm{arcsin} \sqrt{-\omega^\mathrm{SPEP}}\right) &\text{if $\omega^\mathrm{SPEP} < 0$}.
	\end{cases}
\end{equation}
where
\begin{equation} \label{eq:spep_omega}
	\omega^\mathrm{SPEP} \equiv (1 - e^{-\lambda})\left[e^\lambda \bar\rho_a - \bar\rho_b - (e^\lambda - 1)\bar\rho_a\bar\rho_b\right].
\end{equation}
Note that although the result depends explicitly on the sign of $\omega^\mathrm{SPEP}$, it is straightforward to verify that it is an analytic function of $\lambda$. After deriving this result, we compare \eqref{eq:spep_cgf} to the predictions of the hydrodynamic theory. As we show, for large enough currents the two theories, as one might expect using the simple argument of the introduction, do not agree. Finally, we discuss finite-$N$ effects and their implications on the additivity principle.

\subsection{Derivation of the scaled CGF} \label{ssec:spep_cgf}
\label{ssec:ldfSPEPderivation}

As stated above, assuming additivity, the problem of calculating the CGF in the large-$N$ limit is reduced to solving \eqref{eq:time_indep_sol}. To do this it is useful to use the canonical transformation~\cite{DLS2002,Tailleur2007,Tailleur2008}
\begin{equation}  \label{eq:spep_canonical}
	\rho_k = F_k \left[1+(1 - F_k) \hat{F}_k\right], \qquad 	\hat{\rho}_k = \ln \left( 1 + \frac{\hat{F}_k}{1-F_k \hat{F}_k} \right) \;,
\end{equation}
which can also be written as
\begin{equation}
	F_k = \frac{\rho_k}{e^{\hat\rho_k}(1-\rho_k) + \rho_k}\;, 	\qquad 	 \hat{F}_k = (e^{\hat\rho_k}-1)(1-\rho_k) + (1-e^{-\hat\rho_k})\rho_k.
\end{equation}
Then the Hamiltonian in the new set of coordinates, $K^\mathrm{SPEP}_L(\lambda,\bar\rho_a,\bar\rho_b;\mathbf{F},\hat{\mathbf{F}})$, is given by
\begin{align} \label{eq:spep_K}
	K^\mathrm{SPEP}_L(\lambda,\bar\rho_a,\bar\rho_b;\mathbf{F},\hat{\mathbf{F}})
		&= \sum_{k=1}^{L-1} \left[(\hat{F}_{k+1} - \hat{F}_k)(F_k - F_{k+1}) - \hat{F}_k\hat{F}_{k+1}(F_k-F_{k+1})^2\right]
			+ \hat{F}_1 (\bar{\rho}_a - F_1) \nonumber\\
		&\quad + e^{-\lambda}\left[\hat F_L - (e^\lambda - 1)(1-F_L\hat F_L) \right] \left[\bar{\rho}_b - F_L - F_L(1 - \bar{\rho}_b)(e^\lambda - 1)\right].
\end{align}
where $\mathbf{F}=\left(F_1,F_2,\ldots, F_L\right)$ and $\hat{\mathbf{F}}=\left(\hat{F}_1,\hat{F}_2,\ldots, \hat{F}_L\right)$. 

Note that the canonical transformation also adds temporal boundary conditions to the action which can be ignored in the $T \to \infty$ limit. The scaled CGF is then given by:
\begin{align} \label{eq:spep_cgf_K}
	\psi_L(\lambda,\bar\rho_a,\bar\rho_b) = K^\mathrm{SPEP}_L(\lambda,\bar\rho_a,\bar\rho_b;\mathbf{F}^*,\hat{\mathbf{F}}^*),
\end{align}
where $(\mathbf{F}^*,\hat{\mathbf{F}}^*)$ are solutions of
\begin{equation} \label{eq:spep_K_time_indep_sol}
	\frac{\partial \mathbf{F}}{\partial t} = \frac{\partial K^\mathrm{SPEP}_L}{\partial \hat{\mathbf{F}}} = 0,
	\qquad \frac{\partial \hat{\mathbf{F}}}{\partial t} = -\frac{\partial K^\mathrm{SPEP}_L}{\partial \mathbf{F}} = 0.
\end{equation}

In what follows, we solve these equations using the methods used in \cite{Imparato2009}. To avoid cumbersome expressions we drop the $^*$ notation from the optimal profiles $(\mathbf{F}^*,\hat{\mathbf{F}}^*)$ and use $(\mathbf{F},\hat{\mathbf{F}})$. First, we choose the Ansatz
\begin{align} \label{eq:spep_dF_dFh_saddle}
	\hat F_k = -A \sinh (kB), \qquad F_k = \bar\rho_a + \frac{1}{2A}\tanh \frac{kB}{2},
\end{align}
where $A$ and $B$ are undetermined constants. It is easy to check that this Ansatz satisfies \eqref{eq:spep_K_time_indep_sol} for $1 \le k \le L-1$.
Then the constants $A$ and $B$ are determined by the remaining saddle-point equations
\begin{equation}
	\frac{\partial K^\mathrm{SPEP}_L}{\partial \hat F_L} = \frac{\partial K^\mathrm{SPEP}_L}{\partial F_L} = 0.
\end{equation}
These equations imply
\begin{equation}
	\frac{\partial K^\mathrm{SPEP}_L}{\partial F_L} = -4A^2\cosh \frac{LB}{2} \frac{\partial K^\mathrm{SPEP}_L}{\partial \hat F_L},
\end{equation}
from which we obtain
\begin{equation} \label{eq:spep_A_saddle}
	A^2 = \mathcal{A}^\mathrm{SPEP} \equiv \frac{(e^\lambda - 1) [e^\lambda(\bar\rho_b - 1) - \bar\rho_b]}
			{4 [1 + (e^\lambda - 1)\bar\rho_a] [e^\lambda\bar\rho_a(\bar\rho_b-1)-\bar\rho_a\bar\rho_b+\bar\rho_b]} \;.
\end{equation}
Then one can show that $-\frac{1}{A}\frac{\partial K^\mathrm{SPEP}_L}{\partial \hat F_L} = 0$ has the form of
\begin{equation} \label{eq:spep_B_epsilon}
	\sinh(LB - \varepsilon) + \sinh B = 2 \sinh \frac{(L+1)B - \varepsilon}{2} \cosh \frac{(L-1)B - \varepsilon}{2} = 0,
\end{equation}
where $\varepsilon$ satisfies
\begin{equation} \label{eq:spep_omega2}
	\sinh^2 \frac{\varepsilon}{2}
		= \omega^\mathrm{SPEP} \equiv (1 - e^{-\lambda})\left[e^\lambda \bar\rho_a - \bar\rho_b - (e^\lambda - 1)\bar\rho_a\bar\rho_b\right].
\end{equation}
Given $\varepsilon$, \eqref{eq:spep_B_epsilon} is solved by
\begin{equation}
	B = \frac{\varepsilon}{L+1}.
\end{equation}

Thus we have found $A$ and $B$ up to the undetermined signs of $A$ and $\varepsilon$. These signs can be fixed by noting that the optimal density profile $\brho^*$ must always be nonnegative and that the CGF must vanish at $\lambda = 0$. Without loss of generality, for $\bar\rho_a \ge \bar\rho_b$ the optimal profiles are given by
\begin{align} \label{eq:spep_profiles}
	\hat F_k^\mathrm{SPEP} &=
	\begin{cases}
		-\sqrt{\mathcal{A}^\mathrm{SPEP}} \sinh \left( \frac{2k}{L+1} \, \mathrm{arcsinh} \sqrt{\omega^\mathrm{SPEP}}\right)
		&\text{ if $\lambda < -\ln \left[\frac{\bar\rho_a(1-\bar\rho_b)}{\bar\rho_b(1-\bar\rho_a)}\right]$,} \\
		-\sqrt{-\mathcal{A}^\mathrm{SPEP}} \sin \left( \frac{2k}{L+1} \, \mathrm{arcsin} \sqrt{-\omega^\mathrm{SPEP}}\right)
		&\text{  if $-\ln \left[\frac{\bar\rho_a(1-\bar\rho_b)}{\bar\rho_b(1-\bar\rho_a)}\right] \le \lambda < 0$,} \\
		\sqrt{\mathcal{A}^\mathrm{SPEP}} \sinh \left( \frac{2k}{L+1} \, \mathrm{arcsinh} \sqrt{\omega^\mathrm{SPEP}}\right)
		&\text{  if $\lambda \ge 0$.}
	\end{cases} \nonumber \\
	F_k^\mathrm{SPEP} &=
	\begin{cases}
		\bar\rho_a + \frac{1}{2\sqrt{\mathcal{A}^\mathrm{SPEP}}} \tanh \left( \frac{k}{L+1} \, \mathrm{arcsinh} \sqrt{\omega^\mathrm{SPEP}}\right)
		&\text{ if $\lambda < -\ln \left[\frac{\bar\rho_a(1-\bar\rho_b)}{\bar\rho_b(1-\bar\rho_a)}\right]$,} \\
		\bar\rho_a - \frac{1}{2\sqrt{-\mathcal{A}^\mathrm{SPEP}}} \tan \left( \frac{k}{L+1} \, \mathrm{arcsin} \sqrt{-\omega^\mathrm{SPEP}}\right)
		&\text{  if $-\ln \left[\frac{\bar\rho_a(1-\bar\rho_b)}{\bar\rho_b(1-\bar\rho_a)}\right] \le \lambda < 0$,} \\
		\bar\rho_a - \frac{1}{2\sqrt{\mathcal{A}^\mathrm{SPEP}}} \tanh \left( \frac{k}{L+1} \, \mathrm{arcsinh} \sqrt{\omega^\mathrm{SPEP}}\right)
		&\text{  if $\lambda \ge 0$.} \\
	\end{cases} \nonumber \\
\end{align}
We note that $\mathcal{A}^\mathrm{SPEP}$ and $\omega^\mathrm{SPEP}$ are negative for the intermediate range $-\ln \left[\frac{\bar\rho_a(1-\bar\rho_b)}{\bar\rho_b(1-\bar\rho_a)}\right] < \lambda < 0$ and nonnegative otherwise. The results for $\bar\rho_a < \bar\rho_b$ are easily obtained by a sign change $\lambda \to -\lambda$ and an exchange of $\bar\rho_a$ and $\bar\rho_b$. Using these results with \eqref{eq:spep_cgf_K}, after some algebra one obtains \eqref{eq:spep_cgf}.

\subsection{Comparison with hydrodynamic results} \label{ssec:spep_hydro}

\begin{figure}
	\includegraphics[width = 0.7\textwidth]{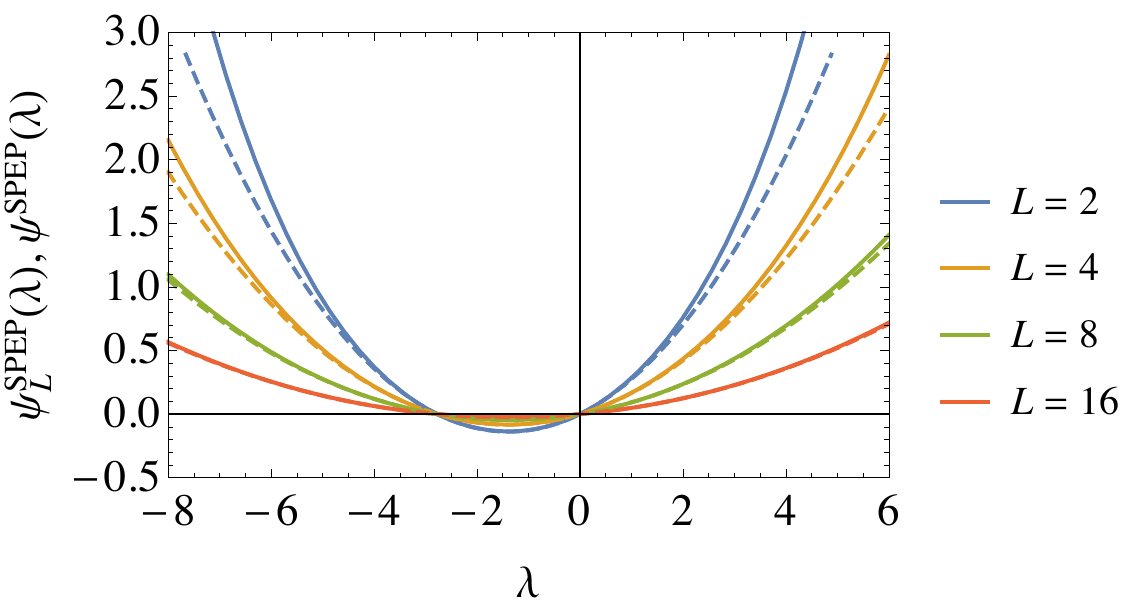}
	\caption{\label{fig:spep_cgf_lambda_psi} The scaled CGFs of the SPEP: the large $N$ limit $\psi_L^\text{SPEP}(\lambda)$ given by \eqref{eq:spep_cgf} (solid lines) and its hydrodynamic limit $\psi^\text{SPEP}(\lambda)$ given by \eqref{eq:spep_cgf_hydro} (dashed lines). For any fixed $\lambda$, $\psi_L^\text{SPEP}(\lambda)$ is equivalent to $\psi^\text{SPEP}(\lambda)$ as $L \to \infty$. The boundary conditions are given by $\bar\rho_a = 0.8$, $\bar\rho_b = 0.2$.}
\end{figure}

We now compare the results of the large-$N$ limit with the predictions of the hydrodynamic theory. The latter has been derived in~\cite{Bodineau2004,Imparato2009} (for the SSEP which shares the same hydrodynamic theory) and can also be obtained by holding $\lambda$ fixed in \eqref{eq:spep_cgf} and taking the large $L$ limit. The expression is given by
\begin{equation} \label{eq:spep_cgf_hydro}
	\psi^\mathrm{SPEP}(\lambda) =
		\begin{cases}
			\frac{1}{L+1} \mathrm{arcsinh}^2 \sqrt{\omega^\mathrm{SPEP}} &\text{if $\omega^\mathrm{SPEP} \ge 0$}, \\
			-\frac{1}{L+1} \mathrm{arcsin}^2 \sqrt{-\omega^\mathrm{SPEP}} &\text{if $\omega^\mathrm{SPEP} < 0$} \;,
		\end{cases}
\end{equation}
and the convergence to it is illustrated in Fig.~\ref{fig:spep_cgf_lambda_psi}. In fact, one can show analytically that
\begin{equation}
	\psi^\mathrm{SPEP}_L(\lambda) - \psi^\mathrm{SPEP}(\lambda)
		= \frac{\mathrm{arcsinh}^4\sqrt{\omega^\mathrm{SPEP}}}{3(L+1)^3} + O\left((L+1)^{-4}\right),
\end{equation}
The sign of the leading correction term indicates that the lattice structure increases the magnitude of the current fluctuations.

\begin{figure}
	\includegraphics[width = 0.7\textwidth]{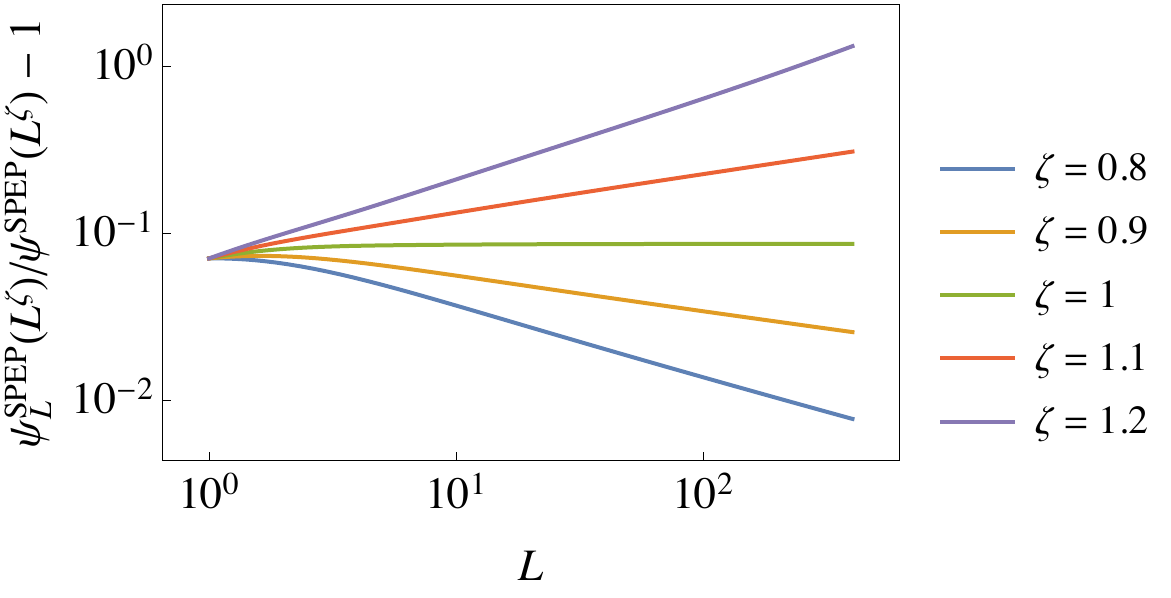}
	\caption{\label{fig:spep_hydro_breakdown} Breakdown of the hydrodynamic limit in the SPEP for $\bar\rho_a = 0.8$ and $\bar\rho_b = 0.2$. (Left) If $\lambda \sim L^\zeta$, $\psi_L$ and $\psi$ are equivalent to each other as $L\to\infty$ only for $\zeta < 1$. (Right) The scaled CGFs of current in the large $N$ limit (solid lines) and in the hydrodynamic limit (dashed lines). If $\lambda = L$, the CGFs do not converge to each other in the $L \to \infty$ limit.}
\end{figure}

To check the validity of the hydrodynamic predictions we next increase $\lambda$ as $\lambda \sim L^\zeta$. This gives
\begin{equation}
	\lim_{L \to \infty} \frac{\psi^\mathrm{SPEP}_L(\lambda)}{\psi^\mathrm{SPEP}(\lambda)} =
		\begin{cases}
			1					&\text{ if $\zeta < 1$,} \\
			\frac{4}{\Lambda^2}\sinh^2 \frac{\Lambda}{2}	&\text{ if $\zeta = 1$ with $\lambda=\Lambda L$} \\
			\infty					&\text{ if $\zeta > 1$.}
		\end{cases}
\end{equation}
This indicates that, as one would naively expect, the hydrodynamic description fails for sufficiently large currents. The threshold separating the hydrodynamic regime from the non-hydrodynamic regime is given by $\lambda \sim L$ (see Fig.~\ref{fig:spep_hydro_breakdown}).

\begin{figure}
	\includegraphics[width = 0.45\textwidth]{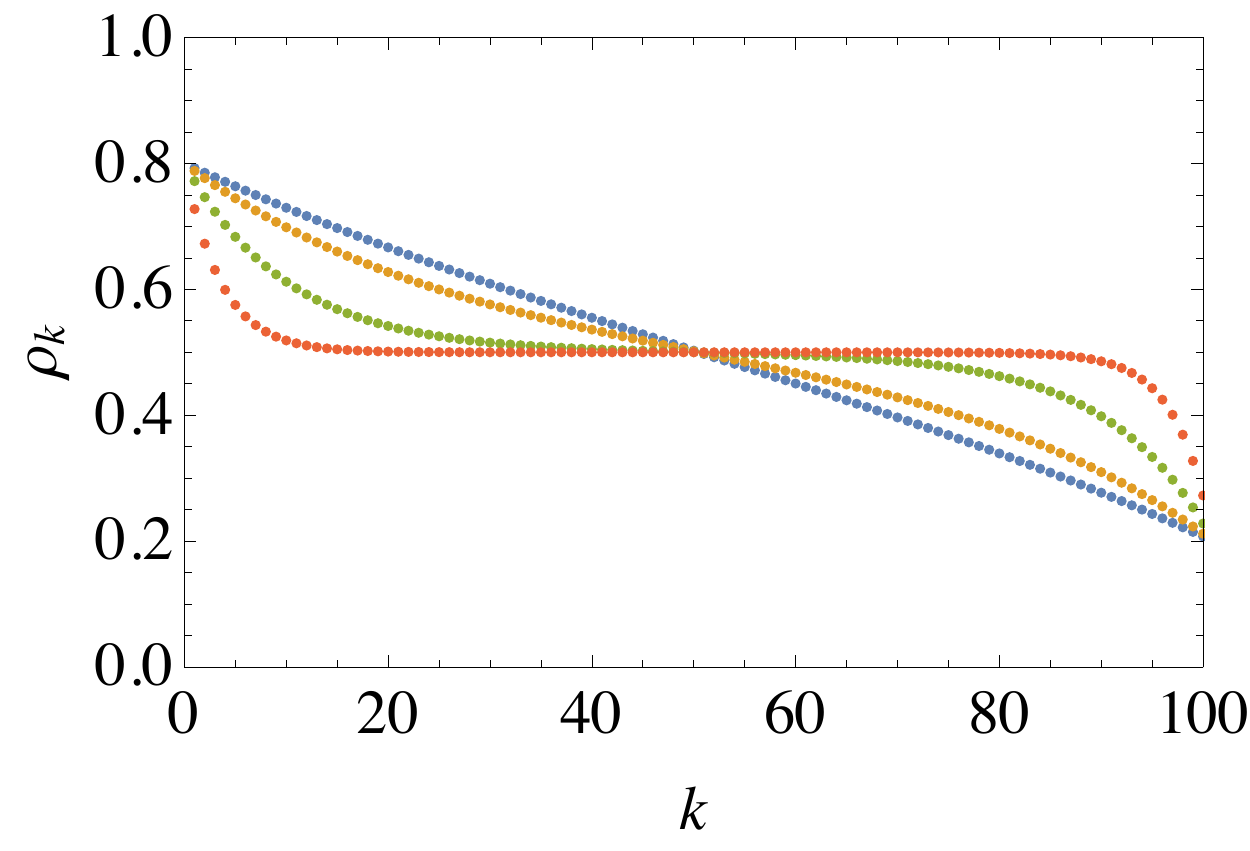}\quad
	\includegraphics[width = 0.45\textwidth]{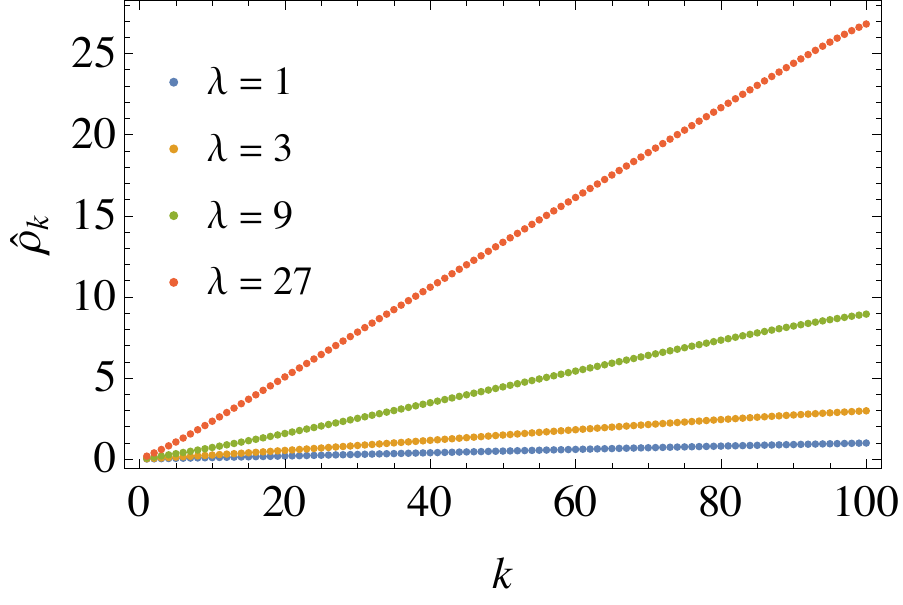}
	\caption{\label{fig:spep_profile} The optimal profiles of the SPEP for $L = 100$, $\bar\rho_a = 0.8$, and $\bar\rho_b = 0.2$, obtained from \eqref{eq:spep_canonical} and~\eqref{eq:spep_profiles}. (Left) As $\lambda$ is increased, the density profile forms a flat bulk profile at $\rho = 1/2$ so that the factor $\rho(1-\rho)$ in $\delta J$ is maximized. (Right) As $\lambda$ is increased, the slope of the momentum profile increases so that its contribution to the time-averaged current becomes larger (see \eqref{eq:spep_J_profiles}).}
\end{figure}

\begin{figure}
	\includegraphics[width = 0.45\textwidth]{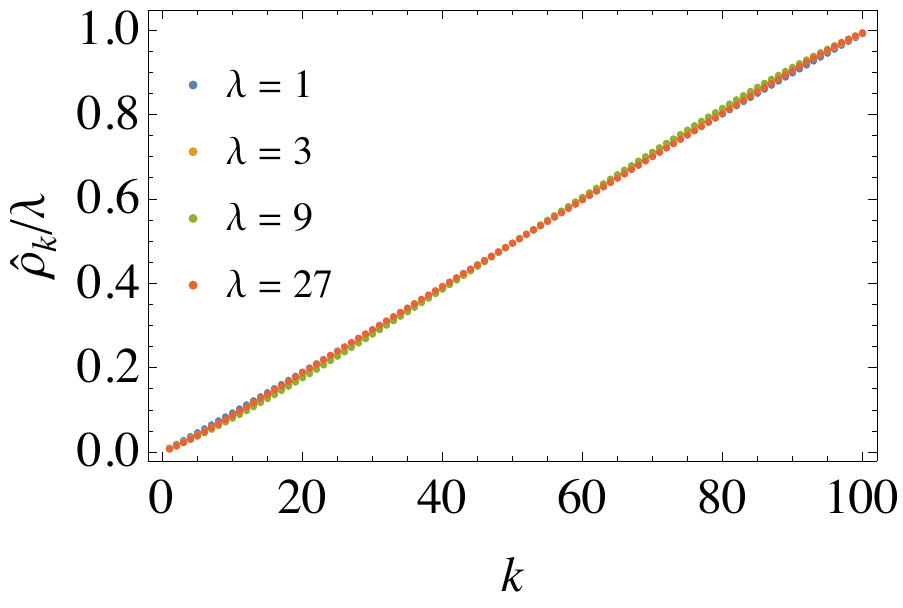}\quad
	\includegraphics[width = 0.45\textwidth]{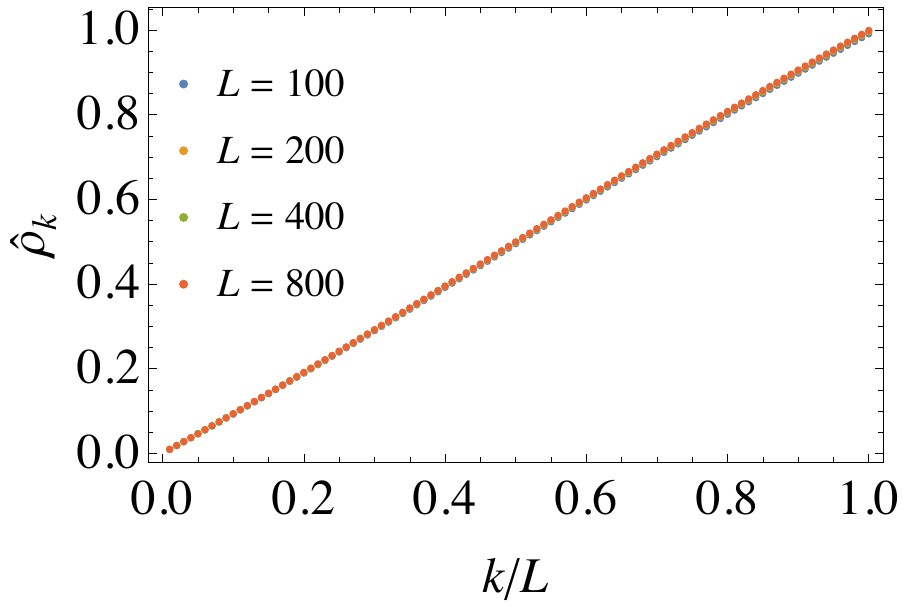}
	\caption{\label{fig:spep_rhoh_collapse} (Left) Momentum profiles at different values of $\lambda$ can be collapsed by using $\hrho_k/\lambda$ as the vertical axis. The other parameters are fixed at $L = 100$, $\bar\rho_a = 0.8$, and $\bar\rho_b = 0.2$. (Right) Momentum profiles at different values of $L$ can be collapsed by using $k/L$ as the horizontal axis. The other parameters are fixed at $\lambda = 1$, $\bar\rho_a = 0.8$, and $\bar\rho_b = 0.2$.}
\end{figure}

As we later show, there are other models where the predictions of the hydrodynamic theory hold well beyond the naive expectation. To this end it is useful to see in detail how the predictions of the hydrodynamic limit fail for the SPEP. To do this, we note that Hamilton's equation takes the form of
\begin{equation}
	\dot{\rho_k} = \frac{\partial H^\mathrm{SPEP}_L}{\partial \hrho_k} = J_{k-1,k} - J_{k,k+1},
\end{equation}
where $J_{k,k+1}$ is the current from box $k$ to box $k+1$. The time-averaged current $J$ can be expressed in terms of the  optimal profiles (again we drop the $^*$ notation) as
\begin{align} \label{eq:spep_J_profiles}
	J &= \frac{1}{L+1}\sum_{k=0}^L\left[(\rho_k - \rho_{k+1})+\rho_k(1-\rho_{k+1})(e^{\hrho_{k+1}-\hrho_k}-1)
		- \rho_{k+1}(1-\rho_k)(e^{\hrho_k-\hrho_{k+1}}-1)\right] \nonumber\\
	&= \underbrace{\frac{\bar\rho_a-\bar\rho_b}{L+1}}_{= \langle J \rangle} + \underbrace{\frac{1}{L+1}
		\sum_{k=0}^L\left[\rho_k(1-\rho_{k+1})(e^{\hrho_{k+1}-\hrho_k}-1) - \rho_{k+1}(1-\rho_k)(e^{\hrho_k-\hrho_{k+1}}-1)\right]}_{= \delta J}.
\end{align}
Since the mean value $\langle J \rangle$ always scales as $1/L$, large values of $J$ are always dominated by the fluctuation $\delta J$ (see \cite{Meerson2014} for a similar observation). Next, note that, as shown in Fig.~\ref{fig:spep_profile}, a large $\delta J$ is supported by a plateau of the density profile close to $\rho = 1/2$ and a  slope of the momentum profile which grows with $\lambda$ (and hence with $J$). In addition, as indicated by the data collapses in Fig.~\ref{fig:spep_rhoh_collapse}, the momentum profile has the scaling form
\begin{equation}
	\hrho_k(\lambda,L) \simeq \lambda g(k/L).
\end{equation}
This implies that
\begin{equation}
	\hrho_{k+1}(\lambda,L) - \hrho_k(\lambda,L) \simeq \frac{\lambda}{L} g'(k/L) \simeq L^{\zeta - 1} \partial_x g.
\end{equation}
If $\zeta < 1$, the momentum gradient decreases with $L$. Then we can approximate $\delta J$ as
\begin{align}\label{eq:spep_J_hydro}
	\delta J \simeq L^{\zeta - 1} \int_0^1 \rmd x\, 2 \rho(1-\rho) \partial_x g,
\end{align}
whose integral form suggests that the current is blind to the lattice structure for any $\zeta < 1$. In other words, the current does not feel any difference between the case $\zeta = 0$ (which can be considered as proper hydrodynamic regime) and the case $0 < \zeta < 1$. Thus its fluctuations show hydrodynamic behaviors in both cases. On the other hand, if $\zeta \ge 1$, the momentum gradient increases with $L$. Then the approximate \eqref{eq:spep_J_hydro} becomes invalid, and the current becomes sensitive to the lattice structure. Thus $\zeta = 1$, which corresponds to $J = O(L^0)$ by \eqref{eq:spep_J_profiles}, is the threshold separating the hydrodynamic regime from the non-hydrodynamic one. We note that this threshold is larger than what one would naively expect from the simple argument given in Sec.~\ref{sec:intro}, {\it i.e.,} $J = O(L^{-1})$.

\subsection{Finite-$N$ corrections and the validity of the additivity principle} \label{ssec:spep_fin_N}

In what follows, we analyze the leading finite-$N$ correction to the scaled CGF $\psi^\mathrm{SPEP}_L$. This provides  a useful tool for numerical corroboration of our analytical results, and confirms the stability of the time-independent saddle-point profiles. The latter thus supports the validity of the additivity principle for the SPEP.

As explained in Appendix~\ref{app:finiteNquantumfluctuationsSPEP}, one can integrate spatio-temporal fluctuations around the saddle-point optimal solutions. This is done by using a mapping (generalizing that of Ref.~\cite{Lecomte2010}) between the CGF of the system with reservoirs at generic densities $\bar\rho_a$, $\bar\rho_b$ and the CGF for reservoirs at densities $\frac 12$.
The resulting expression is finite and analytic, which proves that the additivity hypothesis is correct with respect to continuous phase transitions towards time-dependent profiles (which, if they had existed, would have implied an instability of $\brho^*,\hbrho^*$, reflected in a singularity of the correction). 
%
%
The saddle-point contribution $\psi_L(\lambda)$ to the CGF is complemented by a $1/N$ correction:
\begin{equation} \label{eq:spep_cgf_fin_N_series}
  \psi^\mathrm{SPEP}_{N,L}(\lambda)=\psi^\mathrm{SPEP}_L(\lambda) +N^{-1} \psi_L^{1,\mathrm{SPEP}}(\lambda) + o(N^{-1})
\end{equation}
with, denoting $L'=L+1$,
\begin{align}
\psi_L^{1,\mathrm{SPEP}}(\lambda)
&=
\sum_{p=1}^{L'-1}
\bigg\{
\boldsymbol c_\lambda
-
\cos\frac{p\pi}{2L'}
\sqrt{\frac 12 \boldsymbol c_\lambda\big(\boldsymbol c_\lambda+\cos\frac{p\pi}{L'}\big)}
-
\sin\frac{p\pi}{2L'}
\sqrt{\frac 12 \boldsymbol c_\lambda\big(\boldsymbol c_\lambda-\cos\frac{p\pi}{L'}\big)}
\bigg\}
\label{eq:respsiLhalfdensities_sym}
\end{align}
where we defined $\boldsymbol c_\lambda=  \cosh \frac{2\operatorname{arcsinh}\sqrt{\omega^{\text{SPEP}}}}{L'}$.

\begin{figure}
	\includegraphics[width = 0.6\textwidth]{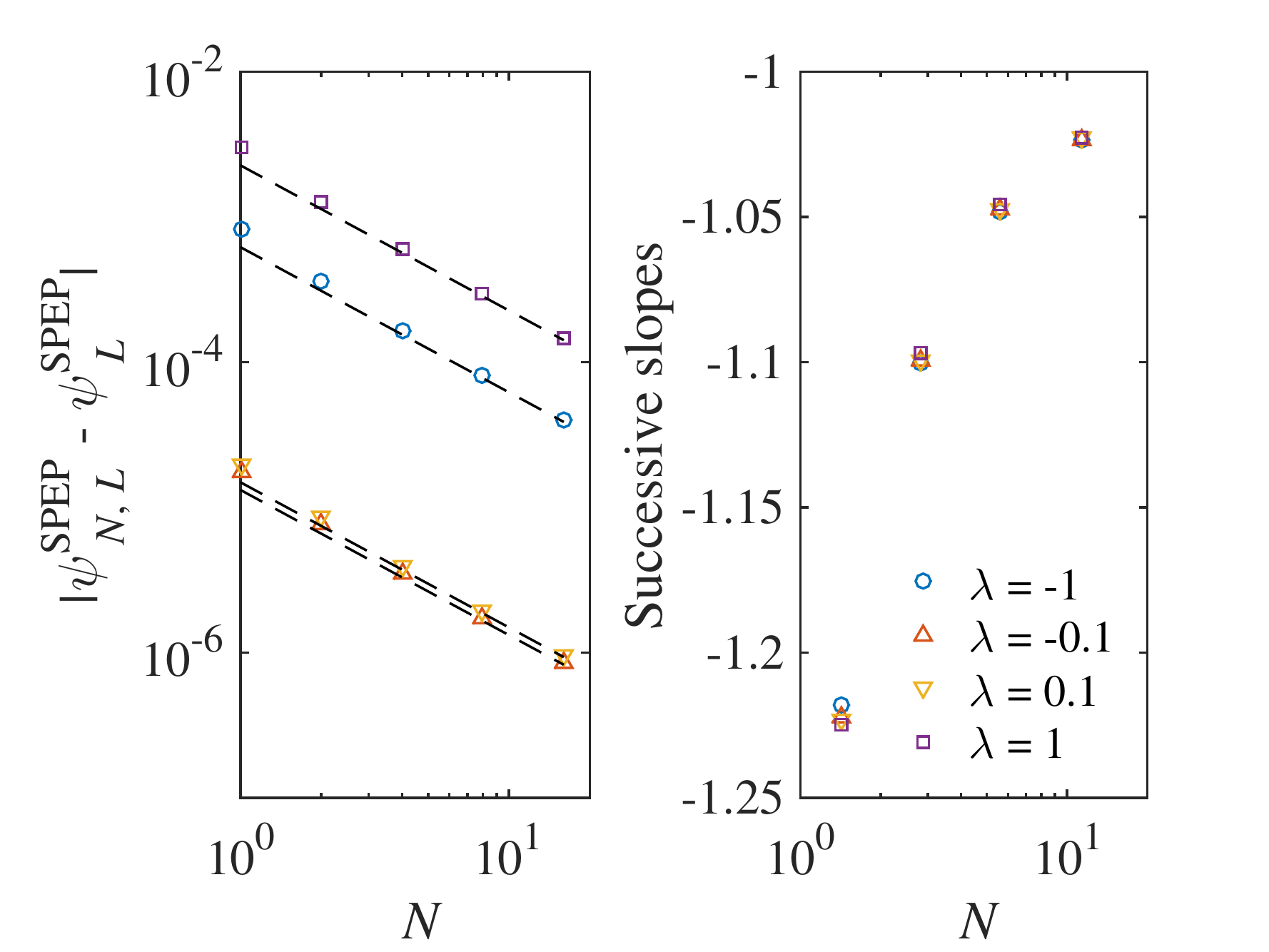}
	\caption{\label{fig:spep_fin_N} Finite-$N$ corrections to the scaled CGF of the SPEP at $L = 3$, $\bar\rho_a = 0.8$, and $\bar\rho_b = 0.2$. The numerics (symbols) are in good agreement with the leading-order correction (dashed lines) predicted by \eqref{eq:respsiLhalfdensities_sym}. The predicted scaling exponent is also supported by the successive slopes of finite-$N$ corrections in the log-log plot.}
\end{figure}

We numerically confirm our theoretical predictions by implementing a finite-$N$ propagator of the SPEP conditioned on a given value of $\lambda$. The eigenvalue with the largest real part corresponds to the scaled CGF $\psi^\mathrm{SPEP}_{N,L}$. As shown in Fig.~\ref{fig:spep_fin_N}, our theory correctly predicts the leading-order behaviors of $\psi^\mathrm{SPEP}_{N,L} - \psi^\mathrm{SPEP}_L$.

We now detail how the large-$L$ limit (at fixed $\lambda$) of \eqref{eq:respsiLhalfdensities_sym} matches the MFT result obtained for the SSEP~\cite{appert-rolland_universal_2008}.
The $L\to\infty$ limit behavior of \eqref{eq:respsiLhalfdensities_sym} is not immediately extractable; following a procedure described in Appendix~\ref{app:finiteNquantumfluctuationsSPEP}, one obtains
\begin{align}
  \psi_L^{1,\mathrm{SPEP}}(\lambda) 
&=
  \frac 1{8L^2} \mathcal F\big(\!-\!\mu(\lambda)\big) \ + \
  O(L^{-3}).
\label{eq:largeLpsi1L}
\end{align}
Here, with $\mu(\lambda)=\operatorname{arcsinh}^2\sqrt{\omega^{\text{SPEP}}}$, we recognize the universal scaling function
\begin{align} \label{eq:univ_scaling}
  \mathcal F(u) 
  &= 4 \sum_{p=1}^{\infty}
  \Big\{
   (p\pi)^2+u-p\pi\sqrt{(p\pi)^2-2u }
   \Big\}
\end{align}
as the one also arising in MFT~\cite{appert-rolland_universal_2008} and Bethe-Ansatz~\cite{appert-rolland_universal_2008,prolhac_cumulants_2009} studies of current fluctuations.
The large-$L$ limit (at fixed $\lambda$) thus yields the same correction as in the MFT approach~\cite{Imparato2009} for the SSEP.
The universal scaling function $\mathcal F(u)$ is singular at a positive value $u_\text{c}=\pi^2/2$ of its argument, but this value is never reached for any real-valued $\lambda$ in \eqref{eq:largeLpsi1L}.
This confirms, as in the MFT context, that the additivity principle holds at large $L$.
%


\section{Current large deviations of SIP}  \label{sec:sip_current}

In this section, we derive the scaled CGF of the time-averaged current of the SIP in the large-$N$ limit. It is given by
\begin{equation} \label{eq:sip_cgf}
	\psi^\mathrm{SIP}_L (\lambda,\bar\rho_a,\bar\rho_b) = \begin{cases}
		(L+1) \sin^2 \left(\frac{1}{L+1} \mathrm{arcsin} \sqrt{\omega^\mathrm{SIP}}\right) &\text{if $\omega^\mathrm{SIP} \ge 0$}, \\
		-(L+1) \sinh^2 \left(\frac{1}{L+1} \mathrm{arcsinh} \sqrt{-\omega^\mathrm{SIP}}\right) &\text{if $\omega^\mathrm{SIP} < 0$}
	\end{cases}
\end{equation}
with the differences between $\mathrm{SIP}(1)$ and $\mathrm{SIP}(1+\alpha)$ encoded in
\begin{equation} \label{eq:sip_omega}
	\omega^\mathrm{SIP} = \begin{cases}
		\omega^{\mathrm{SIP}(1)} \equiv (1 - e^{-\lambda})\left[e^\lambda \bar\rho_a - \bar\rho_b + (e^\lambda - 1)\bar\rho_a\bar\rho_b\right] &\text{ for $\mathrm{SIP}(1)$.} \\
		\omega^{\mathrm{SIP}(1+\alpha)} \equiv \lambda\rho_a - \lambda\rho_b + \lambda^2 \rho_a \rho_b &\text{ for $\mathrm{SIP}(1+\alpha)$.}
	\end{cases}
\end{equation}
Again, one can easily verify that $\psi^\mathrm{SIP}_L$ does not have any singularity at $\omega^\mathrm{SIP} = 0$. A comparison of this result with the hydrodynamic theory shows that for both $\mathrm{SIP}(1)$ and $\mathrm{SIP}(1+\alpha)$ arbitrarily large current fluctuations are still correctly captured by the hydrodynamic theory, in contrast to the SPEP. We close the section with a discussion of finite-$N$ effects.

\subsection{Derivation of the scaled CGF} \label{ssec:sip_cgf}

\subsubsection{$\mathrm{SIP}(1)$}

Similarly to the SPEP, we first transform $H^{\mathrm{SIP}(1)}_L$ into a more convenient form. This is done by the canonical transformation
\begin{equation} \label{eq:sip1_canonical}
	\rho_k = F_k \left[1+(1 + F_k) \hat{F}_k\right], \qquad \hrho_k = \ln \left( 1 + \frac{\hat{F}_k}{1+F_k \hat{F}_k} \right),
\end{equation}
which can also be written as
\begin{equation}
	F_k = \frac{\rho_k}{e^{\hrho_k}(1+\rho_k)-\rho_k}, \qquad \hat{F}_k = (e^{\hrho_k}-1)(1+\rho_k) - (1-e^{-\hrho_k})\rho_k.
\end{equation}
After this transformation, the Hamiltonian of the new variables is given by
\begin{align} \label{eq:sip1_K}
	K^{\mathrm{SIP}(1)}_L(\lambda,\bar\rho_a,\bar\rho_b;\mathbf{F},\hat{\mathbf{F}})
		&= \sum_{k=1}^{L-1} \left[(\hat{F}_{k+1} - \hat{F}_k)(F_k - F_{k+1}) + \hat{F}_k\hat{F}_{k+1}(F_k-F_{k+1})^2\right]
			+ \hat{F}_1 (\bar{\rho}_a - F_1) \nonumber\\
		&\quad + e^{-\lambda}\Big[\hat F_L - (e^\lambda - 1)(1+F_L\hat F_L) \Big] \Big[\bar{\rho}_b - F_L - F_L(1 + \bar{\rho}_b)(e^\lambda - 1)\Big].
\end{align}
Comparing this expression with \eqref{eq:spep_K}, we find the formal correspondence
\begin{equation} \label{eq:sip1_K_spep_K}
	K^{\mathrm{SIP}(1)}_L (\lambda,\bar\rho_a,\bar\rho_b;\mathbf{F},\hat{\mathbf{F}})
		= -K^\mathrm{SPEP}_L (\lambda,-\bar\rho_a,-\bar\rho_b;-\mathbf{F},\hat{\mathbf{F}}).
\end{equation}
By examining the time-independent saddle-point equations derived from these Hamiltonians, we find a mapping between the optimal profiles of the SPEP and the $\mathrm{SIP}(1)$:
\begin{align} \label{eq:sip1_spep_mapping}
	(\hat F^*_k)^{\mathrm{SIP}(1)} (\lambda,\bar\rho_a,\bar\rho_b)	&= (\hat F^*_k)^\mathrm{SPEP} (\lambda,-\bar\rho_a,-\bar\rho_b), \nonumber\\
	(F^*_k)^{\mathrm{SIP}(1)} (\lambda,\bar\rho_a,\bar\rho_b)		&= -(F^*_k)^\mathrm{SPEP} (\lambda,-\bar\rho_a,-\bar\rho_b).
\end{align}

This mapping can be used to obtain the optimal profiles and the scaled CGF of the $\mathrm{SIP}(1)$ from those of the SPEP. It should be noted that, when $\bar\rho_a$ and $\bar\rho_b$ are negative, we should reconsider the proper signs of $A$ and $\varepsilon$ in the optimal profiles of the SPEP. In this case, the optimal density profile $\brho^*$ must always be nonpositive, so that it becomes nonnegative after the mapping to the $\mathrm{SIP}(1)$. Taking this into account, for $\bar\rho_a \ge \bar\rho_b$, the optimal profiles are given by (dropping $^*$)
\begin{align} \label{eq:sip1_profiles}
	\hat F_k^{\mathrm{SIP}(1)} &=
	\begin{cases}
		-\sqrt{\mathcal{A}^{\mathrm{SIP}(1)}} \sin \left( \frac{2k}{L+1} \, \mathrm{arcsin} \sqrt{\omega^{\mathrm{SIP}(1)}}\right)
		&\text{ if $-\ln \left(1+\frac{1}{\rho_b}\right) < \lambda < -\ln \left[\frac{\bar\rho_a(1+\bar\rho_b)}{\bar\rho_b(1+\bar\rho_a)}\right]$,} \\
		-\sqrt{\mathcal{-A}^{\mathrm{SIP}(1)}} \sinh \left( \frac{2k}{L+1} \, \mathrm{arcsinh} \sqrt{-\omega^{\mathrm{SIP}(1)}}\right)
		&\text{ if $-\ln \left[\frac{\bar\rho_a(1+\bar\rho_b)}{\bar\rho_b(1+\bar\rho_a)}\right] \le \lambda < 0$,} \\
		\sqrt{\mathcal{A}^{\mathrm{SIP}(1)}} \sin \left( \frac{2k}{L+1} \, \mathrm{arcsin} \sqrt{\omega^{\mathrm{SIP}(1)}}\right)
		&\text{ if $0 \le \lambda < \ln \left(1 + \frac{1}{\rho_a}\right)$,}
	\end{cases} \nonumber \\
	F_k^{\mathrm{SIP}(1)} &=
	\begin{cases}
		\bar\rho_a + \frac{1}{2\sqrt{\mathcal{A}^{\mathrm{SIP}(1)}}} \tan \left( \frac{k}{L+1} \, \mathrm{arcsin} \sqrt{\omega^{\mathrm{SIP}(1)}}\right)
		&\text{ if $-\ln \left(1+\frac{1}{\rho_b}\right) < \lambda < -\ln \left[\frac{\bar\rho_a(1+\bar\rho_b)}{\bar\rho_b(1+\bar\rho_a)}\right]$,} \\
		\bar\rho_a - \frac{1}{2\sqrt{-\mathcal{A}^{\mathrm{SIP}(1)}}} \tanh \left( \frac{k}{L+1} \, \mathrm{arcsinh} \sqrt{-\omega^{\mathrm{SIP}(1)}}\right)
		&\text{ if $-\ln \left[\frac{\bar\rho_a(1+\bar\rho_b)}{\bar\rho_b(1+\bar\rho_a)}\right] \le \lambda < 0$,} \\
		\bar\rho_a - \frac{1}{2\sqrt{\mathcal{A}^{\mathrm{SIP}(1)}}} \tan \left( \frac{k}{L+1} \, \mathrm{arcsin} \sqrt{\omega^{\mathrm{SIP}(1)}}\right)
		&\text{ if $0 \le \lambda < \ln \left(1 + \frac{1}{\rho_a}\right)$,}
	\end{cases}
\end{align}
where we defined
\begin{align}
	\mathcal{A}^{\mathrm{SIP}(1)} (\lambda,\bar\rho_a,\bar\rho_b)
		&\equiv -\mathcal{A}^\mathrm{SPEP} (\lambda,-\bar\rho_a,-\bar\rho_b) \nonumber\\
		&= \frac{(e^\lambda - 1) [e^\lambda(\bar\rho_b + 1) - \bar\rho_b]}
			{4 [1 - (e^\lambda - 1)\bar\rho_a] [e^\lambda\bar\rho_a(\bar\rho_b+1)-\bar\rho_a\bar\rho_b-\bar\rho_b]}, \label{eq:sip1_A_saddle} \\
	\omega^{\mathrm{SIP}(1)} (\lambda,\bar\rho_a,\bar\rho_b)
		&\equiv -\omega^\mathrm{SPEP} (\lambda,-\bar\rho_a,-\bar\rho_b). \label{eq:sip_omega_mapping-spep}
\end{align}
Note that this definition of $\omega^{\mathrm{SIP}(1)}$ yields \eqref{eq:sip_omega}. The expressions for $\bar\rho_a < \bar\rho_b$ are obtained by $\lambda \to -\lambda$ and an exchange of $\bar\rho_a$ and $\bar\rho_b$.

Finally, due to \eqref{eq:sip1_K_spep_K} and \eqref{eq:sip1_spep_mapping}, the scaled CGFs of the SPEP and the $\mathrm{SIP}(1)$ are related by
\begin{equation}
	\psi^{\mathrm{SIP}(1)}_L (\lambda,\bar\rho_a,\bar\rho_b) = -\psi^\mathrm{SPEP}_L (\lambda,-\bar\rho_a,-\bar\rho_b)\;,
\end{equation}
from which it is straightforward to derive \eqref{eq:sip_cgf}.

Remarkably, the $\mathrm{SIP}(1)$ has a finite range of $\lambda$, whereas the SPEP has an unbounded range of $\lambda$. This is related to the fact that the domain of $\mathrm{arcsin}$ is limited to $[-1,1]$, while that of $\mathrm{arcsinh}$ is unlimited. As will be discussed later, the limited range of $\lambda$ is closely related to the persistence of hydrodynamic behaviors for extreme current fluctuations. Meanwhile, it should be noted that the limited range of $\lambda$ does not imply a limited range of the current being considered. The time-averaged current $J$ conditioned on $\lambda$, obtained from $\partial \psi^{\mathrm{SIP}(1)}_L/\partial \lambda$, still ranges from $-\infty$ to $\infty$ for both SPEP and $\mathrm{SIP}(1)$. In fact, using standard Legendre transform arguments it is easy to check that the limited range of definition of the CGF $\psi^{\mathrm{SIP}(1)}_L(\lambda)$ corresponds to exponential tails of the current distribution function. 

\subsubsection{$\mathrm{SIP}(1+\alpha)$}

We now turn to the case of $\mathrm{SIP}(1+\alpha)$. Again, the Hamiltonian, given by \eqref{eq:sip1a_H}, can be simplified by a canonical transformation
\begin{equation} \label{eq:sip1a_canonical}
	\rho_k = F_k (1+F_k\hat{F}_k), \qquad \hrho_k = \frac{\hat{F}_k}{1+F_k \hat{F}_k} \\
\end{equation}
or
\begin{equation}
	F_k = \frac{\rho_k}{1+\rho_k\hrho_k}, \quad \hat{F}_k = \hrho_k(1+\rho_k\hrho_k),
\end{equation}
which was also used in \cite{Tailleur2008} in the context of the equivalent KMP model. This transforms the Hamiltonian into
\begin{align}\label{eq:sip1a_K}
	K^{\mathrm{SIP}(1+\alpha)}_L(\lambda,\bar\rho_a,\bar\rho_b;\mathbf{F},\hat{\mathbf{F}})
		&= \sum_{k=1}^{L-1} \left[(\hat{F}_{k+1} - \hat{F}_k)(F_k - F_{k+1})
			+ \hat{F}_k\hat{F}_{k+1}(F_k-F_{k+1})^2\right] + \hat{F}_1 (\bar{\rho}_a - F_1) \nonumber\\
		&\quad + \Big[\hat F_L - \lambda(1+F_L\hat F_L) \Big] \Big[\bar{\rho}_b - F_L - \bar{\rho}_b\lambda F_L\Big].
\end{align}
A comparison between this expression and \eqref{eq:sip1_K} shows
\begin{equation} \label{eq:sip1a_K_sip1_K}
	K^{\mathrm{SIP}(1+\alpha)}_L (\lambda,\bar\rho_a,\bar\rho_b;\mathbf{F},\hat{\mathbf{F}})
		= \lim_{N \to \infty} K^{\mathrm{SIP}(1)}_L
			(N^{-\alpha}\lambda,N^\alpha\bar\rho_a,N^\alpha\bar\rho_b;N^\alpha\mathbf{F},N^{-\alpha}\hat{\mathbf{F}}).
\end{equation}
The time-independent saddle-point equations of these Hamiltonians show that the optimal profiles of the $\mathrm{SIP}(1)$ and the $\mathrm{SIP}(1+\alpha)$ are related by (again dropping $^*$)
\begin{align} \label{eq:sip1a_sip1_mapping}
	(\hat F_k)^{\mathrm{SIP}(1+\alpha)} (\lambda,\bar\rho_a,\bar\rho_b)
		&= \lim_{N \to \infty} N^\alpha (\hat F_k)^{\mathrm{SIP}(1)} (N^{-\alpha}\lambda,N^\alpha\bar\rho_a,N^\alpha\bar\rho_b), \nonumber\\
	(F_k)^{\mathrm{SIP}(1+\alpha)} (\lambda,\bar\rho_a,\bar\rho_b)
		&= \lim_{N \to \infty} N^{-\alpha} (F_k)^{\mathrm{SIP}(1)} (N^{-\alpha}\lambda,N^\alpha\bar\rho_a,N^\alpha\bar\rho_b).
\end{align}
Therefore, for $\bar\rho_a \ge \bar\rho_b$ the optimal profiles are obtained as 
\begin{align} \label{eq:sip1a_profiles}
	\hat F_k^{\mathrm{SIP}(1+\alpha)} &=
	\begin{cases}
		-\sqrt{\mathcal{A}^{\mathrm{SIP}(1+\alpha)}} \sin \left( \frac{2k}{L+1} \, \mathrm{arcsin} \sqrt{\omega^{\mathrm{SIP}(1+\alpha)}}\right)
		&\text{ if $-\frac{1}{\bar\rho_b} < \lambda < \frac{1}{\bar\rho_a} - \frac{1}{\bar\rho_b}$,} \\
		-\sqrt{\mathcal{-A}^{\mathrm{SIP}(1+\alpha)}} \sinh \left( \frac{2k}{L+1} \, \mathrm{arcsinh} \sqrt{-\omega^{\mathrm{SIP}(1+\alpha)}}\right)
		&\text{ if $\frac{1}{\bar\rho_a} - \frac{1}{\bar\rho_b} \le \lambda < 0$,} \\
		\sqrt{\mathcal{A}^{\mathrm{SIP}(1+\alpha)}} \sin \left( \frac{2k}{L+1} \, \mathrm{arcsin} \sqrt{\omega^{\mathrm{SIP}(1+\alpha)}}\right)
		&\text{ if $0 \le \lambda < \frac{1}{\bar\rho_a}$,}
	\end{cases} \nonumber \\
	F_k^{\mathrm{SIP}(1+\alpha)} &=
	\begin{cases}
		\bar\rho_a + \frac{1}{2\sqrt{\mathcal{A}^{\mathrm{SIP}(1+\alpha)}}} \tan \left( \frac{k}{L+1} \, \mathrm{arcsin} \sqrt{\omega^{\mathrm{SIP}(1+\alpha)}}\right)
		&\text{ if $-\frac{1}{\bar\rho_b} < \lambda < \frac{1}{\bar\rho_a} - \frac{1}{\bar\rho_b}$,} \\
		\bar\rho_a - \frac{1}{2\sqrt{-\mathcal{A}^{\mathrm{SIP}(1+\alpha)}}} \tanh \left( \frac{k}{L+1} \, \mathrm{arcsinh} \sqrt{-\omega^{\mathrm{SIP}(1+\alpha)}}\right)
		&\text{ if $\frac{1}{\bar\rho_a} - \frac{1}{\bar\rho_b} \le \lambda < 0$,} \\
		\bar\rho_a - \frac{1}{2\sqrt{\mathcal{A}^{\mathrm{SIP}(1+\alpha)}}} \tan \left( \frac{k}{L+1} \, \mathrm{arcsin} \sqrt{\omega^{\mathrm{SIP}(1+\alpha)}}\right)
		&\text{ if $0 \le \lambda < \frac{1}{\bar\rho_a}$,}
	\end{cases}
\end{align}
where we defined
\begin{align}
	\mathcal{A}^{\mathrm{SIP}(1+\alpha)} (\lambda,\bar\rho_a,\bar\rho_b)
		&\equiv \lim_{N \to \infty} N^{2\alpha}\mathcal{A}^{\mathrm{SIP}(1)}
			(N^{-\alpha}\lambda,N^\alpha\bar\rho_a,N^\alpha\bar\rho_b) \nonumber\\
		&= \frac{\lambda(1+\lambda\rho_b)}{4 (1-\lambda\rho_a) (\rho_a - \rho_b + \lambda \rho_a \rho_b)}, \label{eq:sip1a_A_saddle} \\
	\omega^{\mathrm{SIP}(1+\alpha)}
		&\equiv \lim_{N \to \infty} \omega^{\mathrm{SIP}(1)} (N^{-\alpha}\lambda,N^\alpha\bar\rho_a,N^\alpha\bar\rho_b).
\end{align}
Note that this definition of $\omega^{\mathrm{SIP}(1+\alpha)}$ leads to \eqref{eq:sip_omega}.

Using \eqref{eq:sip1a_K_sip1_K} and \eqref{eq:sip1a_sip1_mapping}, the scaled CGFs of the $\mathrm{SIP}(1)$ and the $\mathrm{SIP}(1+\alpha)$ are related by
\begin{equation}
	\psi^{\mathrm{SIP}(1+\alpha)}_L (\lambda,\bar\rho_a,\bar\rho_b)
		= \lim_{N \to \infty} \psi^{\mathrm{SIP}(1)}_L (N^{-\alpha}\lambda,N^\alpha\bar\rho_a,N^\alpha\bar\rho_b),
\end{equation}
which gives \eqref{eq:sip_cgf}. As in the case of the $\mathrm{SIP}(1)$, $\mathrm{SIP}(1+\alpha)$ also has a finite range of $\lambda$, although the range of the current $J$ is unbounded.

\subsection{Comparison with hydrodynamic results} \label{ssec:sip_hydro}

\begin{figure}
	\includegraphics[width = 0.45\textwidth]{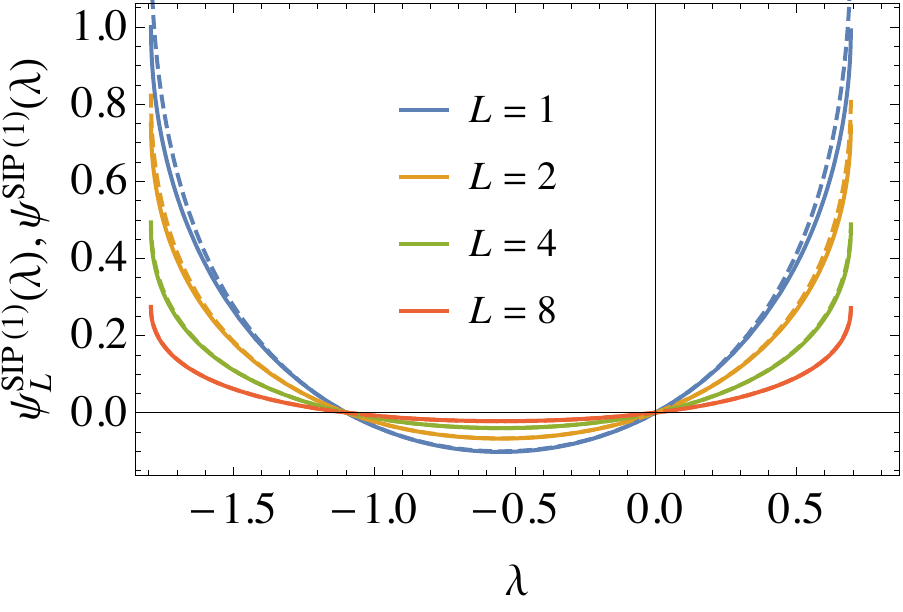} \quad
	\includegraphics[width = 0.45\textwidth]{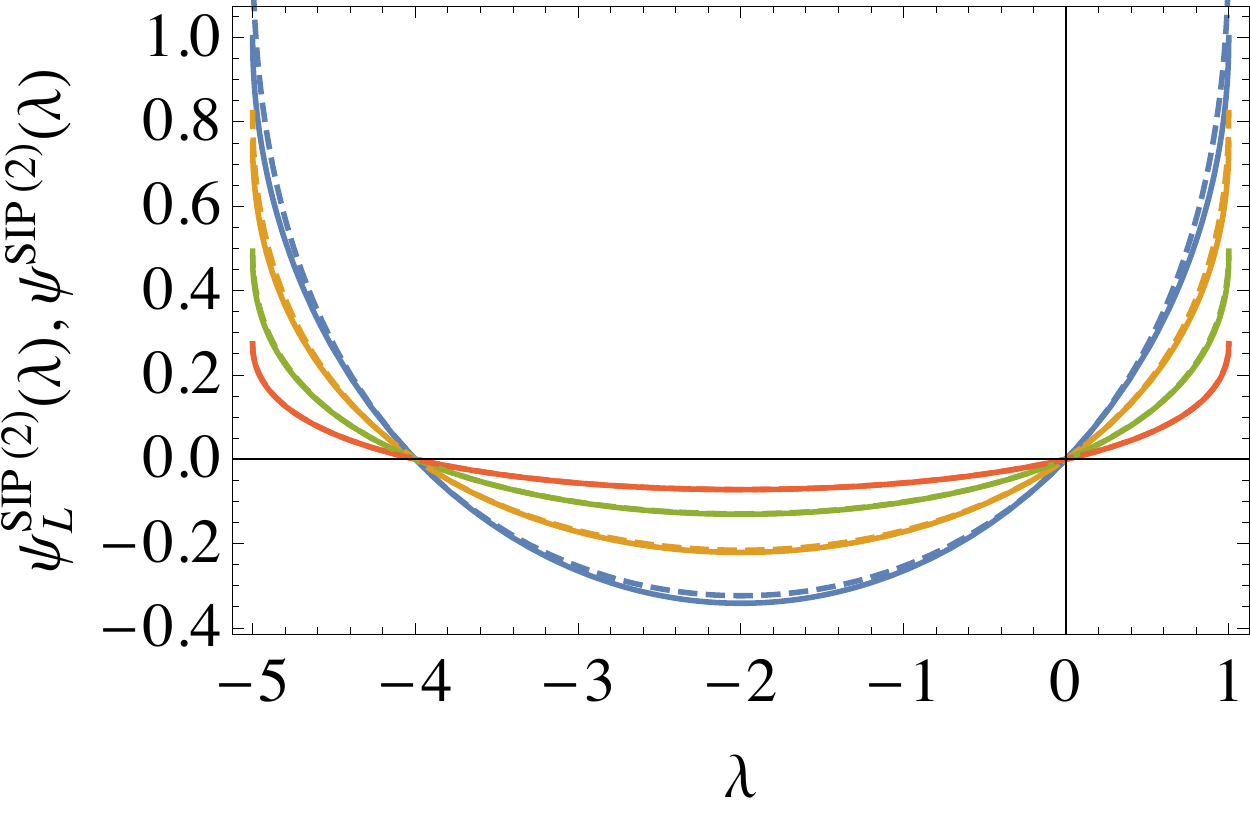}
	\caption{\label{fig:sip_cgf} The scaled CGFs of (left) $\mathrm{SIP}(1)$ and (right) $\mathrm{SIP}(2)$ in the large $N$ limit ($\psi^\mathrm{SIP}_L (\lambda)$, \eqref{eq:sip_cgf}, solid lines) and in the hydrodynamic limit ($\psi^\mathrm{SIP} (\lambda)$, \eqref{eq:sip_cgf_hydro}, dashed lines). For any fixed $\lambda$, the CGF $\psi^\mathrm{SIP}_L (\lambda)$ is equivalent to $\psi^\mathrm{SIP} (\lambda)$ as $L \to \infty$. The boundary conditions are given by $\bar\rho_a = 0.8$, $\bar\rho_b = 0.2$.}
\end{figure}

For both $\mathrm{SIP}(1)$ and $\mathrm{SIP}(1+\alpha)$, in the $L \to \infty$ limit, the hydrodynamic expression of the scaled CGF can be written in a similar form
\begin{equation} \label{eq:sip_cgf_hydro}
	\psi^\mathrm{SIP}(\lambda) = \begin{cases}
		\frac{1}{L+1} \mathrm{arcsin}^2 \sqrt{\omega^\mathrm{SIP}} &\text{if $\omega^\mathrm{SIP} \ge 0$}, \\
		-\frac{1}{L+1} \mathrm{arcsinh}^2 \sqrt{-\omega^\mathrm{SIP}} &\text{if $\omega^\mathrm{SIP} < 0$},
\end{cases}
\end{equation}
where we used the superscript $\mathrm{SIP}$ to refer to both large-$N$ models. This expression is in agreement with the corresponding expression for the KMP model found in \cite{Imparato2009}. When $\lambda$ is fixed, the hydrodynamic limit is reached by
\begin{equation}
	\psi^\mathrm{SIP}_L(\lambda) - \psi^\mathrm{SIP}(\lambda) = -\frac{\mathrm{arcsin}^4\sqrt{\omega^\mathrm{SIP}}}{3L^3} + O(L^{-4}),
\end{equation}
as illustrated in Fig.~\ref{fig:sip_cgf}. In contrast to the SPEP, the lattice structure decreases the magnitude of fluctuations. Since the range of $\lambda$ is bounded and the two CGFs converge to each other throughout this range, we cannot find any scaling of $\lambda$ with $L$ that induces non-hydrodynamic current fluctuations.

\begin{figure}
	\includegraphics[width = 0.45\textwidth]{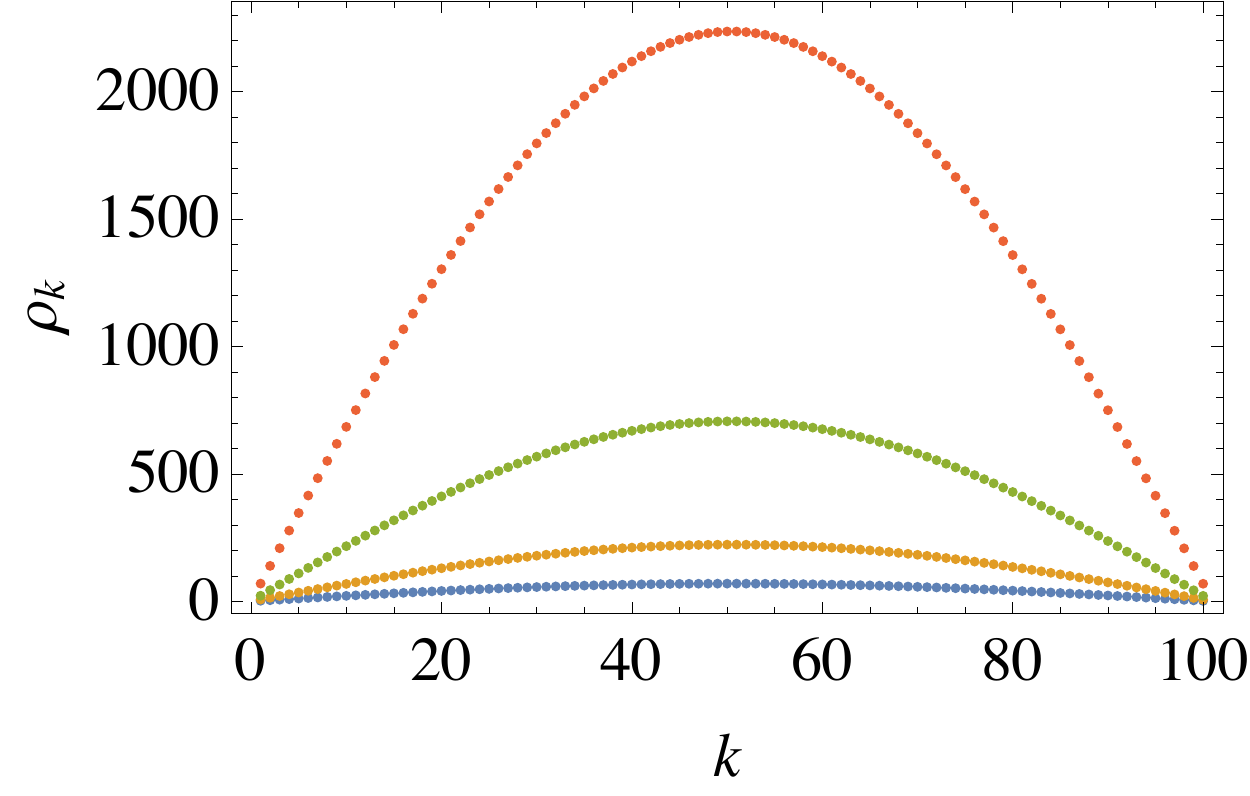}\quad
	\includegraphics[width = 0.45\textwidth]{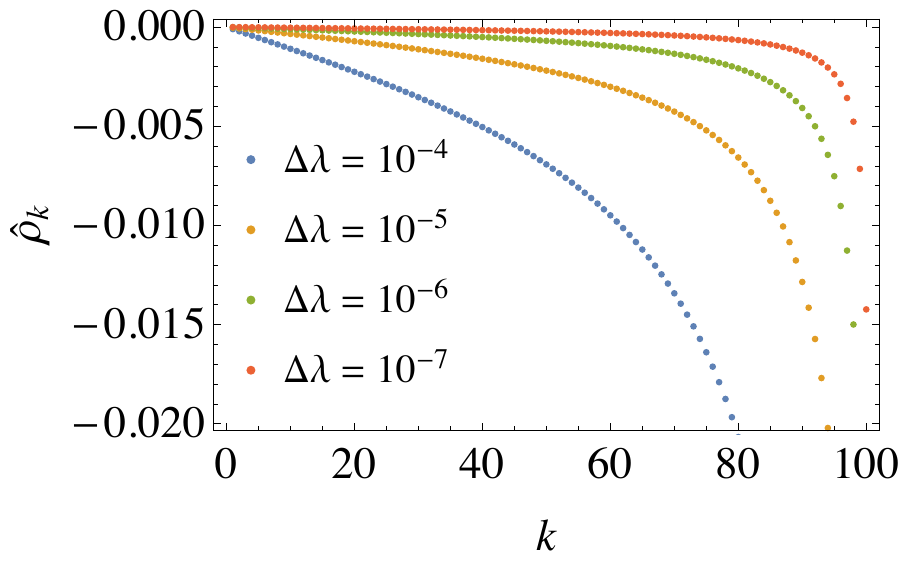}
	\caption{\label{fig:sip1_profile} The optimal profiles of the $\mathrm{SIP}(1)$ for $L = 100$, $\bar\rho_a = 0.8$, $\bar\rho_b = 0.2$. As $\Delta\lambda = \lambda - \lambda_\mathrm{min}$ approaches zero, (left) the density profile develops an arbitrarily large crest, while (right) the momentum profile becomes flatter.}
\end{figure}

\begin{figure}
	\includegraphics[width = 0.45\textwidth]{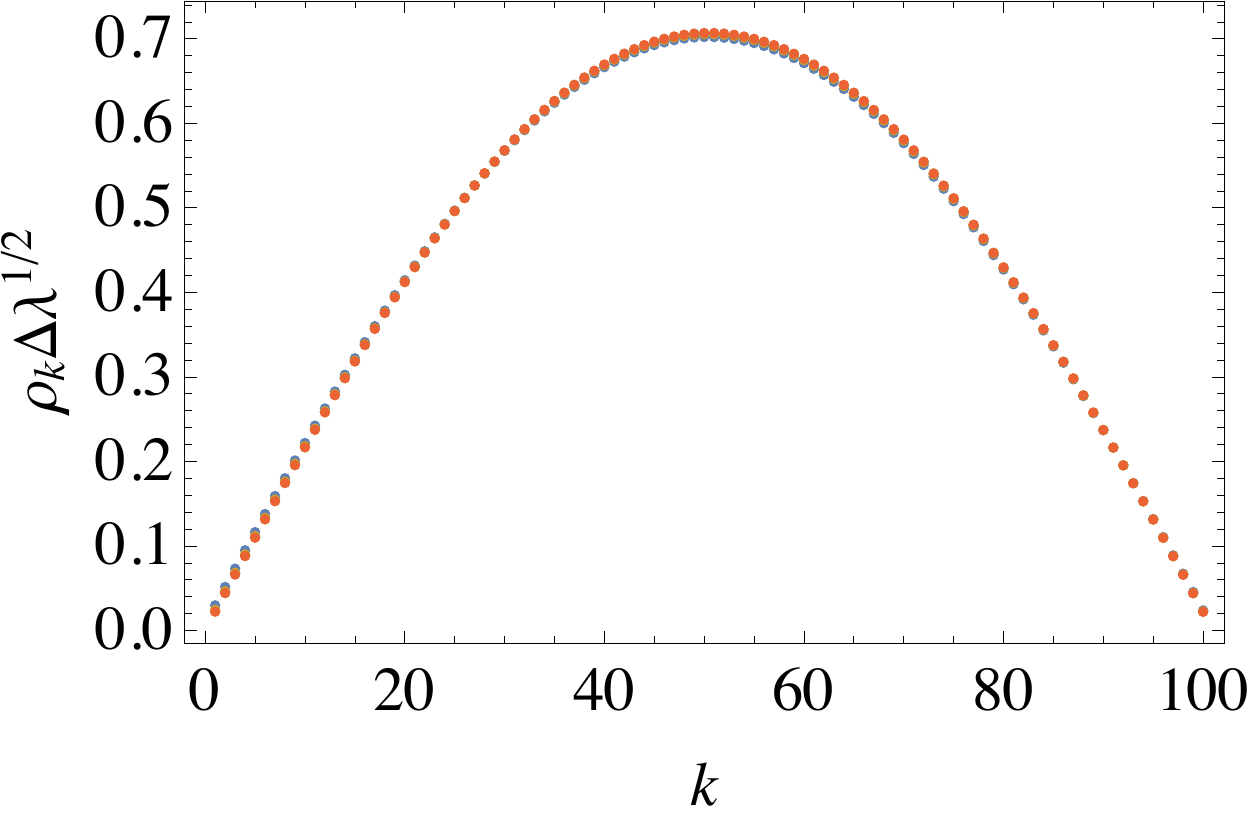}\quad
	\includegraphics[width = 0.45\textwidth]{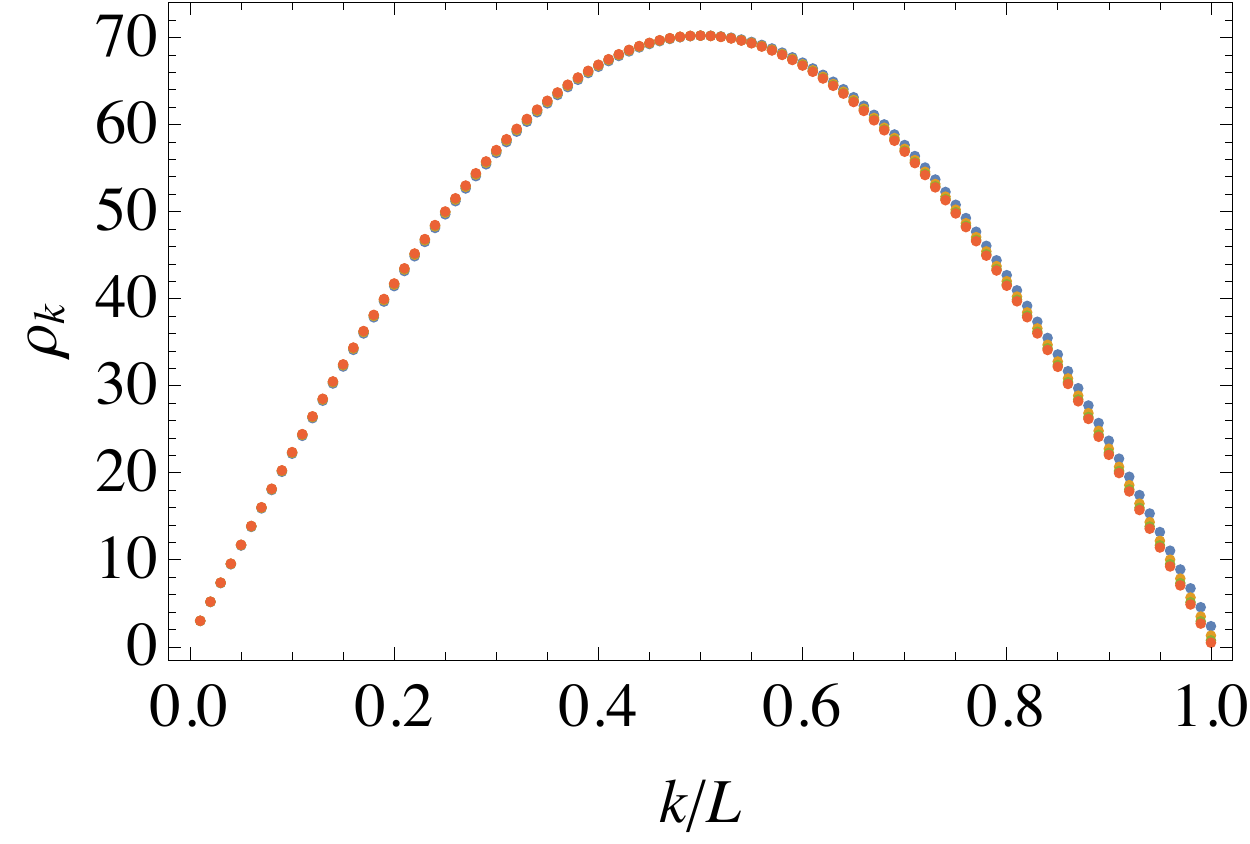}\\
	\includegraphics[width = 0.45\textwidth]{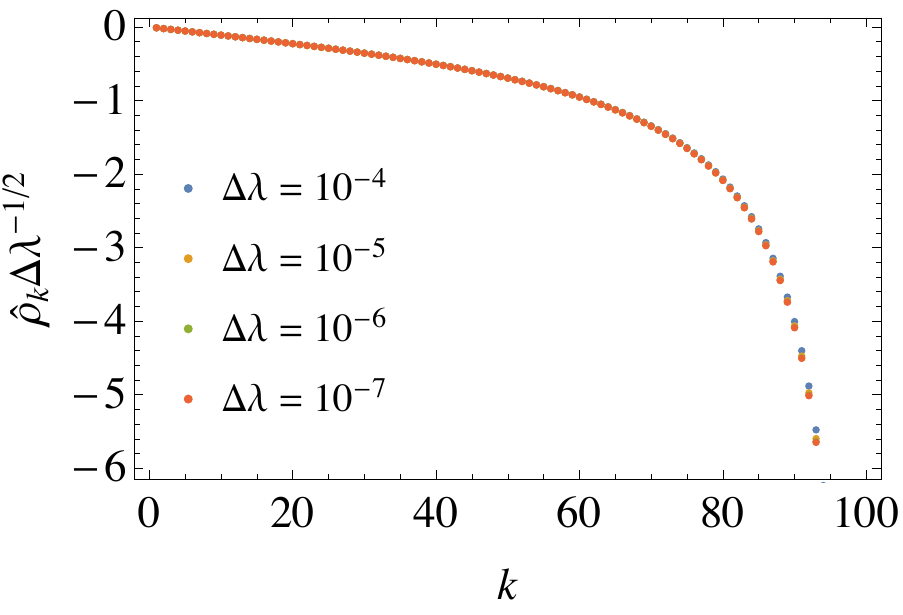}\quad
	\includegraphics[width = 0.45\textwidth]{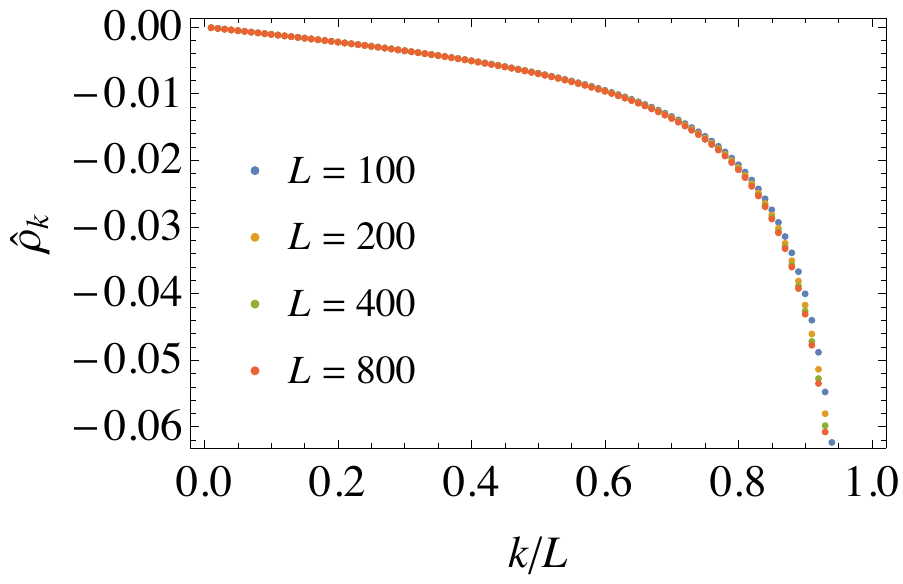}
	\caption{\label{fig:sip1_profile_collapse} The data collapses of optimal profiles of the $\mathrm{SIP}(1)$. The boundary conditions are given by $\bar\rho_a = 0.8$ and $\bar\rho_b = 0.2$. We fix $L = 100$ ($\lambda = 1$) while $\lambda$ ($L$) is varied.}
\end{figure}

%

In order to understand why hydrodynamic behaviors are still observed for arbitrarily large current fluctuations, we examine the optimal profiles of the $\mathrm{SIP}(1)$ as was done for the SPEP. Using the same argument applied to the SPEP, the time-averaged current $J$ of $\mathrm{SIP}(1)$ can be related to the optimal profiles by
\begin{align} \label{eq:sip1_J_profiles}
	J &= \frac{1}{L+1}\sum_{k=0}^L\left[(\rho_k - \rho_{k+1})+\rho_k(1+\rho_{k+1})(e^{\hrho_{k+1}-\hrho_k}-1)
		- \rho_{k+1}(1+\rho_k)(e^{\hrho_k-\hrho_{k+1}}-1)\right] \nonumber\\
	&= \underbrace{\frac{\bar\rho_a-\bar\rho_b}{L+1}}_{= \langle J \rangle} + \underbrace{\frac{1}{L+1}
		\sum_{k=0}^L\left[\rho_k(1+\rho_{k+1})(e^{\hrho_{k+1}-\hrho_k}-1) - \rho_{k+1}(1+\rho_k)(e^{\hrho_k-\hrho_{k+1}}-1)\right]}_{= \delta J},
\end{align}
where $\delta J$ becomes dominant as $\lambda$ approaches its upper and lower bounds
\begin{equation}
	\lambda_\mathrm{max} = \ln \left(1 + \frac{1}{\rho_a}\right), \quad \lambda_\mathrm{min} = -\ln \left(1+\frac{1}{\rho_b}\right).
\end{equation}
For convenience, let us denote by $\Delta \lambda$ both $|\lambda - \lambda_\mathrm{min}|$ and $|\lambda - \lambda_\mathrm{max}|$. As shown in Fig.~\ref{fig:sip1_profile}, a large $\delta J$ is supported by a growing density crest and a flattening momentum profile as $\Delta \lambda \to 0$. As the data collapses in Fig.~\ref{fig:sip1_profile_collapse} indicate, as $L \to \infty$, the optimal profiles have scaling forms
\begin{align} \label{eq:sip1_profile_scaling}
	\rho_k(\Delta\lambda,L) &= \Delta\lambda^{-1/2} f(k/L), \nonumber\\
	\hrho_k(\Delta\lambda,L) &= \Delta\lambda^{1/2} g(k/L).
\end{align}
These imply that we can approximate $\delta J$ as
\begin{align}\label{eq:sip1_J_hydro}
	\delta J &\simeq \frac{1}{L+1}\sum_{k=0}^L[\rho_k(1+\rho_{k+1}) + \rho_{k+1}(1+\rho_k)](\hrho_{k+1}-\hrho_k) \nonumber\\
		&\simeq \frac{1}{\Delta\lambda^{1/2}L} \int_0^1 \rmd x\, 2 \rho(1+\rho) \partial_x g,
\end{align}
which has an integral form for any small $\Delta \lambda$ corresponding to large $J$. Thus, $J$ exhibits hydrodynamic behaviors for arbitrarily large $J$. An almost identical argument also applies to the $\mathrm{SIP}(1+\alpha)$, whose optimal profiles have similar shapes and satisfy the scaling relation \eqref{eq:sip1_profile_scaling}.

\subsection{Finite-$N$ effects} \label{ssec:sip_fin_N}
\subsubsection{$\mathrm{SIP}(1)$}
From \eqref{eq:spep_H} and \eqref{eq:sip1_H}, we observe that the Hamiltonians of the SPEP and the $\mathrm{SIP}(1)$ are related by
\begin{equation}
	H^{\mathrm{SIP}(1)}_L (\lambda,\bar\rho_a,\bar\rho_b;\brho,\hbrho) = -H^\mathrm{SPEP}_L(\lambda,-\bar\rho_a,-\bar\rho_b;-\brho,\hbrho).
\end{equation}
This suggests that the leading finite-$N$ correction to the scaled CGF $\psi^{\mathrm{SIP}(1)}_L$ can be obtained by a Gaussian approximation very similarly to the one applied to the SPEP in Sec.~\ref{ssec:spep_fin_N}. Consequently, the leading finite-$N$ correction is described by analogs of \eqref{eq:spep_cgf_fin_N_series} and \eqref{eq:respsiLhalfdensities_sym}, namely
\begin{equation}
	\psi^{\mathrm{SIP}(1)}_{N,L}(\lambda) = \psi^{\mathrm{SIP}(1)}_L(\lambda) + N^{-1} \psi^{1,\mathrm{SIP}(1)}_L(\lambda) + o(N^{-1})
\end{equation}
with
\begin{align}
\psi^{1,\mathrm{SIP}(1)}_L(\lambda,\bar\rho_a,\bar\rho_b) &= -\psi^{1,\mathrm{SPEP}}_L(\lambda,-\bar\rho_a,-\bar\rho_b) \nonumber\\
&=
-\sum_{p=1}^{L'-1}
\bigg\{
\boldsymbol c_\lambda
-
\cos\frac{p\pi}{2L'}
\sqrt{\frac 12 \boldsymbol c_\lambda\big(\boldsymbol c_\lambda+\cos\frac{p\pi}{L'}\big)}
-
\sin\frac{p\pi}{2L'}
\sqrt{\frac 12 \boldsymbol c_\lambda\big(\boldsymbol c_\lambda-\cos\frac{p\pi}{L'}\big)}
\bigg\},
\label{eq:respsiLhalfdensities_sym_SIP}
\end{align}
where $\boldsymbol c_\lambda =  \cos \frac{2\operatorname{arcsin}\sqrt{\omega^{\text{SIP}(1)}}}{L+1}$.
The correction is again an analytic function of $\lambda$ within its domain, which proves the validity of the additivity principle with respect to continuous transitions, without ruling out possible discontinuous ones.

In the large-$L$ limit, using \eqref{eq:largeLpsi1L} and the first equality of \eqref{eq:respsiLhalfdensities_sym_SIP}, we can also write
\begin{align}
  \psi_L^{1,\mathrm{SIP}(1)}(\lambda) 
&=
  -\frac 1{8L^2} \mathcal F\big(\nu(\lambda)\big) \ + \
  O(L^{-3}),
\label{eq:sip_largeLpsi1L}
\end{align}
with $\mathcal F(u)$ the universal scaling function defined in \eqref{eq:univ_scaling} and $\nu(\lambda)=\operatorname{arcsin}^2\sqrt{\omega^{\text{SIP}(1)}}$. While $\mathcal F(u)$ is singular at $u_\text{c}=\pi^2/2$, $\nu(\lambda)$ cannot be greater than $\pi^2/4$ for any real-valued $\lambda$ in \eqref{eq:sip_largeLpsi1L}. This also confirms the validity of the additivity principle with respect to continuous transitions at large $L$.

\subsubsection{$\mathrm{SIP}(1+\alpha)$}

Unlike the previous models, the leading finite-$N$ correction to $\psi^{\mathrm{SIP}(1+\alpha)}_L$ comes from a different origin. For this model, if we keep the leading finite-$N$ correction, the path integral in \eqref{eq:macro_path_integ} can be rewritten as
\begin{equation}
	e^{NT\psi^{\mathrm{SIP}(1+\alpha)}_{N,L}}
		= \int \D\brho \D\hbrho \, \exp\left\{-N\int_{0}^{T} \mathrm{d}t \,
			\left[ \hbrho\cdot\dot{\brho} - H^{\mathrm{SIP}(1+\alpha)}_L - N^{-\alpha}V^{\mathrm{SIP}(1+\alpha)}_L \right]\right\},
\end{equation}
where
\begin{equation} \label{eq:sip1a_V_rho}
	V^{\mathrm{SIP}(1+\alpha)}_L(\lambda,\bar\rho_a,\bar\rho_b;\brho,\hbrho)
		= \sum_{k=1}^{L-1} \frac{\rho_k + \rho_{k+1}}{2}\,(\hrho_{k+1} - \hrho_k)^2
			+ \frac{\bar\rho_a + \rho_1}{2}\,\hrho_1^2 + \frac{\bar\rho_b + \rho_L}{2}\,(\hrho_L - \lambda)^2.
\end{equation}
Applying a saddle-point approximation as before, we obtain
\begin{equation}
	\psi^{\mathrm{SIP}(1+\alpha)}_{N,L}(\lambda) = \psi^{\mathrm{SIP}(1+\alpha)}_L(\lambda) + N^{-\alpha} \psi^{1,\mathrm{SIP}(1+\alpha)}_L(\lambda) + o(N^{-\alpha})
\end{equation}
with
\begin{equation} \label{eq:sip1a_saddle_rho}
	\psi^{1,\mathrm{SIP}(1+\alpha)}_L(\lambda) = V^{\mathrm{SIP}(1+\alpha)}_L(\lambda;\brho^*,\hbrho^*),
\end{equation}
where $\brho^*$ and $\hbrho^*$ are the optimal profiles determined in Sec.~\ref{ssec:sip_cgf}.

\subsubsection{Numerical results}

\begin{figure}
	\includegraphics[width = 0.49\textwidth]{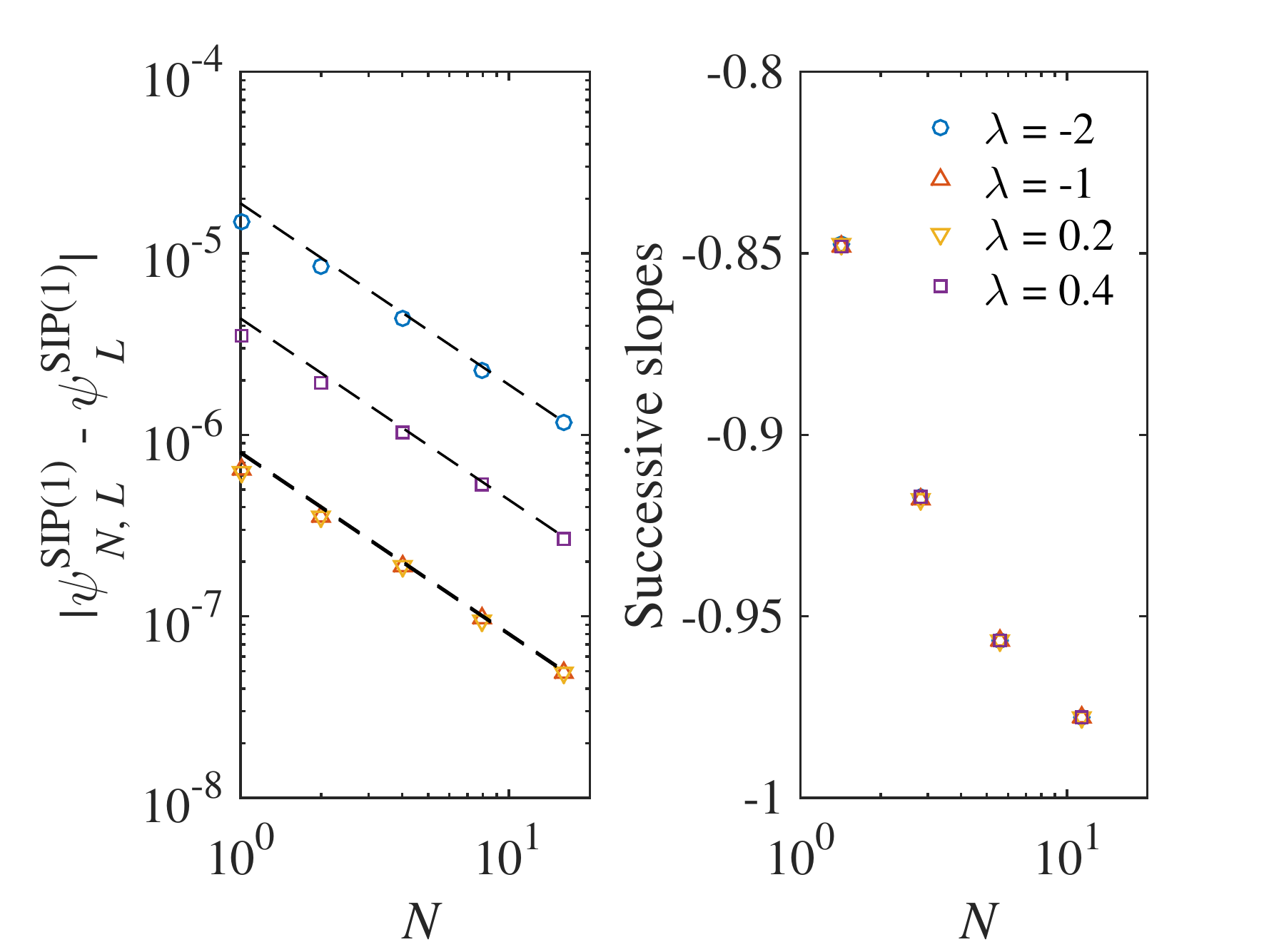}
	\includegraphics[width = 0.49\textwidth]{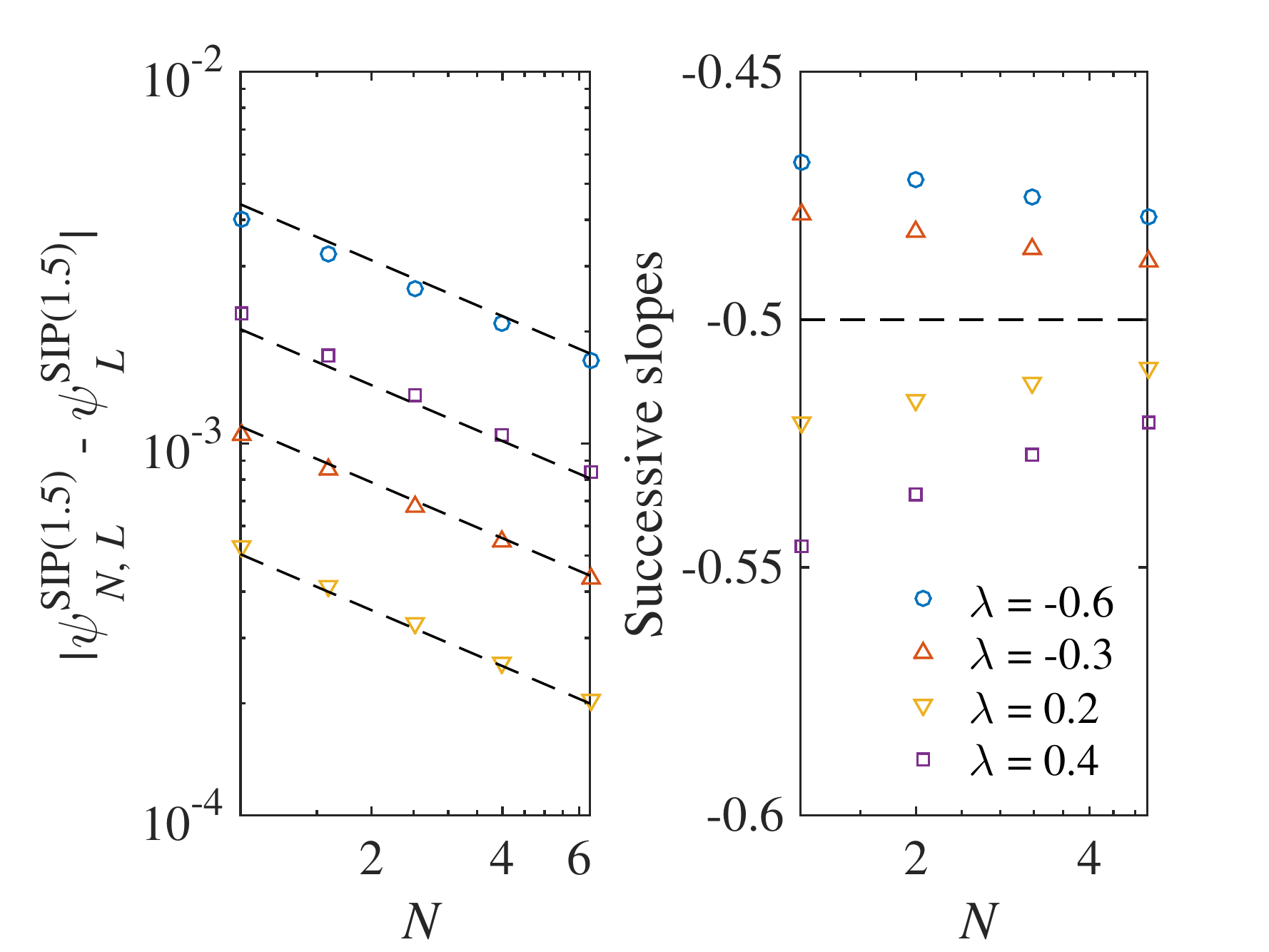}
	\caption{\label{fig:sip_finN} Finite-$N$ corrections to the scaled CGF of (left) the $\mathrm{SIP}(1)$ and (right) the $\mathrm{SIP}(1.5)$ at $L = 3$, $\bar\rho_a = 0.08$, and $\bar\rho_b = 0.02$. The upper cutoff $M$ is set equal to $16$. The numerics (symbols) are in good agreement with the leading-order corrections (dashed lines) given by \eqref{eq:respsiLhalfdensities_sym_SIP} and \eqref{eq:sip1a_saddle_rho}. The predicted scaling exponents are also supported by the successive slopes of finite-$N$ corrections in the log-log plots.}
\end{figure}

We numerically confirm our theoretical predictions by constructing a matrix representation of the SIP conditioned on $\lambda$. Since it is impossible to implement the unbounded configuration space of this model, we introduce an artificial upper bound $M$ on the number of particles in each site. The matrix representation is such that any transition that violates this upper bound is forbidden, while the other transitions occur with the same rates as the original dynamics. We expect that if $M$ is sufficiently large, the effects of $M$ become irrelevant. The results for $\mathrm{SIP}(1)$ and $\mathrm{SIP}(1.5)$ shown in Fig.~\ref{fig:sip_finN} are both in agreement with our predictions.

\section{Criterion for persistent hydrodynamic behaviors} \label{sec:criterion}

We have shown that current fluctuations of the SPEP have a non-hydrodynamic regime, while those of the SIP always behave according to the predictions of the hydrodynamic equations. As noted in Sec.~\ref{sec:models}, one important difference between the SPEP and the SIP lies in whether the mobility coefficient $\sigma(\rho)$ is bounded from above. This suggests a connection between the presence of an upper bound on $\sigma(\rho)$ and hydrodynamic behaviors of current fluctuations. In order to investigate this connection, we examine how the optimal profiles depend on the time-averaged current $J$ within the naive hydrodynamic regime given by $J = O(L^{-1})$. An extrapolation of this dependence beyond the regime ({\em i.e.}, $J$ larger than $O(L^{-1})$) reveals whether non-hydrodynamic behaviors appear for sufficiently large $J$.

In the hydrodynamic limit, from Hamilton's equations we have
\begin{equation} \label{eq:hamilton_hydro}
	\frac{\partial \rho}{\partial t} = \frac{\delta H}{\delta \rho} = \partial_x \left[ D(\rho)\partial_x \rho - \sigma(\rho)\partial_x \hrho \right]
\end{equation}
with $H$ given by \eqref{eq:H_mft}. This gives a relation between $J$ and the optimal profiles through
\begin{align} \label{eq:J_profiles_hydro}
	J = \frac{1}{L+1} \int_0^1 \mathrm{d}x\, \left[-D(\rho^*)\partial_x \rho^* + \sigma(\rho^*)\partial_x \hrho^*\right]
	= \frac{G(\bar\rho_a) - G(\bar\rho_b)}{L+1} + \frac{1}{L+1} \int_0^1 \mathrm{d}x\, \left[\sigma(\rho^*)\partial_x \hrho^*\right],
\end{align}
where $D(\rho) = G'(\rho)$. Then, as long as $G(\bar\rho_a)$ and $G(\bar\rho_b)$ are finite, $J$ beyond the naive hydrodynamic regime satisfies
\begin{align} \label{eq:J_profiles_hydro_approx}
	J \simeq \frac{1}{L+1} \int_0^1 \mathrm{d}x\, \left[\sigma(\rho^*)\partial_x \hrho^*\right].
\end{align}
In other words, in this regime $J$ is sensitive only to $\sigma(\rho^*)$ and $\partial_x \hrho^*$.

Note that when $\sigma(\rho)$ is bounded from above, an arbitrarily large $J$ can only be supported by an arbitrarily large $\partial_x \hrho^*$. This means that $\partial_x \hrho^*$ can no longer be expressed as a proper gradient for a sufficiently large $J$, in which case $J$ exhibits non-hydrodynamic behaviors, as was the case for the SPEP. Hence, the absence of an upper bound on $\sigma(\rho)$ is clearly a necessary condition for the persistence of hydrodynamic behaviors.

When $\sigma(\rho)$ is not bounded from above, a large $J$ can be supported by a large $\sigma(\rho^*)$ while $\partial_x \hrho^*$ remains well defined, so that $J$ is still blind to the lattice structure. Based on this possible scenario, we  conjecture that the absence of an upper bound on $\sigma(\rho)$ is also a sufficient condition for the persistence of hydrodynamic behaviors. Although there is no rigorous proof yet, we can confirm this conjecture for the one-dimensional symmetric zero-range process, which provides a simple example of boundary-driven systems with non-constant $D(\rho)$ and unbounded $\sigma(\rho)$. An interested reader is referred to Appendix~\ref{app:zrp} for more discussions on this model.

\section{Conclusions} \label{sec:conclusions}

In this paper we introduced a class of large-$N$ models for one-dimensional boundary-driven diffusive systems. Using $N$ as a large parameter, we were able to obtain exact expressions for current large deviations on a finite lattice, without relying on a hydrodynamic approach. This allowed us to look at regimes where the hydrodynamic theory is naively expected to break down. Surprisingly, we found that there are classes of models, which we conjecture to be those with an unbounded $\sigma(\rho)$ as a function of $\rho$, where the predictions of the hydrodynamic theory always hold. It will be interesting to see if similar considerations also hold for models with a bulk bias and/or for large deviations of other additive observables, such as the activity.

In addition, we examined the finite-$N$ corrections and used them to argue that the additivity principle, assumed throughout the paper, is likely to hold for the models considered. 

{\it Acknowledgments:} We are grateful for discussions with B.~Derrida, M.~R.~Evans, B.~Meerson, T.~Sadhu, and H.~Spohn. YB and YK were supported by an Israeli-Science-Foundation grant. YB is supported in part at the Technion by a fellowship from the Lady Davis Foundation. VL wishes to thank the hospitality of the Physics Department of Technion, Haifa, where part of the research was performed, and acknowledges support from LAABS Inphyniti CNRS project.

\appendix

\section{Current fluctuations in the hydrodynamic limit} \label{app:hydro_limit}

The limited range of current fluctuations in the hydrodynamic limit, that we discuss in the Introduction, can also be seen from an argument more directly based on the MFT. The scaled CGF $\psi(\lambda)$ for current fluctuations is defined by
\begin{equation}
e^{T_\mathrm{micro}\psi(\lambda)} = \llangle e^{\lambda T_\mathrm{micro} J} \rrangle,
\end{equation}
where $J$ is the mean current across a certain cross-section of the system averaged over a microscopic time interval $t \in [0, T_\mathrm{micro}]$. In the hydrodynamic limit, this expression can be written in a path integral form
\begin{equation}
\llangle e^{\lambda T_\mathrm{micro} J} \rrangle = \int \D\rho \D\hrho \, e^{-\ell \int_0^{T} \rmd t\, \mathcal{L}[\lambda;\rho,\hrho]},
\end{equation}
where $T = T_\mathrm{micro}/\ell^2$ is the length of the time interval on a macroscopic scale. For $\ell \gg 1$, we can apply a saddle-point approximation to obtain
\begin{equation}
\psi(\lambda) = -\frac{1}{T\ell} \inf_{\rho,\hrho} \int_0^{T} \rmd t\, \mathcal{L}[\lambda;\rho,\hrho].
\end{equation}
The minimum action always has the form
\begin{equation}
\inf_{\rho,\hrho} \int_0^{T} \rmd t\, \mathcal{L}[\lambda;\rho,\hrho] = -T f(\lambda),
\end{equation}
from which we obtain
\begin{equation}
\psi(\lambda) = \ell^{-1} f(\lambda).
\end{equation}
Note that the Lagrange multiplier $\lambda$ and the conjugate current $J$ are related by
\begin{equation}
J = \psi'(\lambda) = \ell^{-1} f'(\lambda).
\end{equation}
Thus we recover the conclusion that $O(\ell^{-1})$ current fluctuations belong to the hydrodynamic regime.

\section{Path Integral representation of the CGF} \label{app:path}

These statistics of the current are encoded in the scaled cumulant generating function (CGF) $\psi_L$, which is defined by
\begin{equation} \label{eq:cgf_def_app}
	e^{NT\psi_L(\lambda,\bar\rho_a,\bar\rho_b)} = \llangle e^{N\lambda T J}\rrangle
		= \llangle e^{N^{-\alpha}\lambda \sum_s I_{L,L+1}(t_s)}\rrangle.
\end{equation}
Note that in the last step we divided the time interval $[0,\,T]$ into $M$ infinitesimal subintervals of length $\Delta t$, so that $t_s = s\Delta t$ for $s = 1,\,2,\,\ldots,\,M$, and $T = M\Delta t$. We also introduced the notation
\begin{equation}
	I_{k,k+1}(t_s) = \begin{cases}
		+1 &\text{if a particle hops from $k$ to $k+1$ at $t \in [t_s,t_{s+1}]$}\\
		-1 &\text{if a particle hops from $k+1$ to $k$ at $t \in [t_s,t_{s+1}]$}\\
		0 &\text{otherwise}
	\end{cases}
\end{equation}
for $0 \le s \le M-1$.

From \eqref{eq:cgf_def_app}, the scaled CGF can be expressed in a path integral form
\begin{align} \label{eq:cgf_path_integ}
	&e^{NT\psi_L(\lambda)} = \llangle\int \prod_{s,\,k} \left[\rmd \rho_k(t_s) \,
		\delta\left(\rho_k(t_{s+1}) - \rho_k(t_{s}) - \frac{I_{k-1,k}(t_s) - I_{k,k+1}(t_s)}{N^{1+\alpha}}\right)\right]
		e^{N^{-\alpha} \lambda I_{L,L+1}(t_s)}\rrangle \nonumber \\
	&~ = \int \prod_{s} \llangle\prod_{k = 1}^L\left[\rmd \rho_k(t_s) \rmd \hat \rho_k(t_s) \,
		e^{-N\hat \rho_k(t_s)\left(\rho_k(t_{s+1}) - \rho_k(t_{s}) - \frac{I_{k-1,k}(t_s) - I_{k,k+1}(t_s)}{N^{1+\alpha}}\right)}\right]
		e^{N^{-\alpha}\lambda I_{L,L+1}(t_s)}\rrangle_{\mathbf{I}(t_s)},
\end{align}
which has the standard Martin--Siggia--Rose (MSR) form~\cite{Martin1973,*Janssen1976,*DeDominicis1976,*DeDominicis1978} with auxiliary field variables $\hrho_1, \hrho_2, \ldots, \hrho_L$.

$\mathbf{I}(t_s) = \left(I_{0,1}(t_s), \ldots, I_{k,k+1}(t_s), \ldots, I_{L,L+1}(t_s)\right)$ represents a hopping event within the time interval $[t_s, t_{s+1}]$. The probability distribution of $\mathbf{I}(t_s)$ is given by
\begin{align}
	\mathbf{I}(t_s) = \begin{cases}
		\left(0, \ldots, 0, I_{k,k+1} = 1, 0, \ldots, 0 \right) &\text{with prob. $n_k (N \mp n_{k+1}) \frac{\Delta t}{N}$}, \\
		\left( 0, \ldots, 0, I_{k,k+1} = -1,0, \ldots,0 \right) &\text{with prob. $n_{k+1} (N \mp n_k) \frac{\Delta t}{N}$}, \\
		\left(0,0,\ldots,0\right) &\text{with prob. $1 - \sum_k \left[N(n_k + n_{k+1}) \mp 2n_k n_{k+1}\right] \frac{\Delta t}{N}$},
	\end{cases}
\end{align}
where $k$ is any integer between $0$ and $L+1$, and the upper (lower) rates correspond to the SPEP (SIP). Choosing an appropriate value of the scaling exponent $\alpha$, we can evaluate the average $\langle \cdot \rangle_{\mathbf{I}(t_s)}$ (see \cite{Lefevre2007} for a general description of this procedure) and single out the leading-order component to obtain
\begin{equation} \label{eq:macro_path_integ_app}
	e^{NT\psi_L(\lambda)} = \int \D\brho \D\hbrho \,
		\exp\left\{-N\int_{0}^{T} \mathrm{d}t \,\left[ \hbrho\cdot\dot{\brho}
			- H_L(\lambda;\brho,\hbrho) \right]\right\},
\end{equation}
where the function $H_L$ contains all information about the dynamics. 
Note that the same result can also be derived by a different path-integral construction using $\mathrm{SU}(2)$ (for SPEP) or $\mathrm{SU}(1,1)$ (for SIP) coherent states, extending the one proposed in~\cite{Tailleur2008} to the case of current large deviations.

\section{Finite-$N$ corrections to the CGF for the SPEP arising from space-time fluctuations}
\label{app:finiteNquantumfluctuationsSPEP}
%

In this Appendix, we derive the leading finite-$N$ corrections to the CGF for the SPEP that we obtained in Sec.~\ref{ssec:ldfSPEPderivation} by a large-$N$ saddle-point approach. 
This analysis generalizes the MFT results~\cite{Imparato2009} to the case of the lattice SPEP with finite $L$ and also allows us to discard the existence of a continuous phase transition in the CGF as $\lambda$ is varied, thus (partially) supporting the validity of the additivity principle for the SPEP. To avoid cumbersome expressions, in the following we drop the superscript SPEP.

\subsection{Mapping to reservoirs at half densities}
Determining the finite-$N$ corrections in principle amounts to integrating the quadratic fluctuations around the saddle-point solutions shown in \eqref{eq:spep_profiles}.
This is however rendered difficult by the nontrivial dependence of those solutions on the spatial index $k$ of the lattice.
To bypass this issue, we generalize the approach presented in~\cite{Lecomte2010}:
we map the CGF (taken at $\lambda$) of a system in contact with reservoirs at densities $\bar\rho_a$ and $\bar\rho_b$ to the CGF of a system in contact with reservoirs at same densities $\frac 12$, but taken at a different value of $\lambda$:
\begin{equation}
  \psi_{N,L}(\lambda,\bar\rho_a,\bar\rho_b) = 
  \psi_{N,L}\big(\lambda=2\operatorname{arcsinh} \sqrt{\omega},\bar\rho_a=\tfrac12,\bar\rho_b=\tfrac12\big) 
\label{eq:mmappingpsiLfiniteN}
\end{equation}
This result arises from the $\mathrm{SU}(2)$ symmetry of the generating operator, whose eigenvalue of maximal real part yields the CGF:
(\emph{i}) the bulk part of this operator is left invariant by a $\mathrm{SU}(2)$ rotation as in~\cite{Lecomte2010}, but with spin $\frac N2$ instead of spin $\frac 12$ ;
(\emph{ii}) the terms describing the contact with reservoirs are affected by the rotation and yield \eqref{eq:mmappingpsiLfiniteN} for a well-chosen rotation.

The main advantage of this transformation is that at half densities $\bar\rho_a=\bar\rho_b=\tfrac12$, the saddle-point solutions shown in \eqref{eq:spep_profiles} take a simple form: one has
\begin{equation}
  \hat F_k^*=\sinh\frac{\lambda k}{L+1},
\qquad
  F_k^*=\frac 12-\frac 12\tanh \frac{\lambda k}{2(L+1)}.
\end{equation}
In terms of the original variables $\brho^*$ and $\hbrho^*$, the canonical transformation of \eqref{eq:spep_canonical} gives
\begin{equation}
  \rho_k^*=\frac 12,
\qquad
  \hat\rho_k^*=\lambda\frac{k}{L+1},
\label{eq:rhohatrhoksaddle}
\end{equation}
which shows that the optimal density profile is flat, while the optimal momentum profile is linear. 
We note that the same behavior was also observed in the hydrodynamic limit~\cite{Bodineau2004}.

\subsection{Small space-time fluctuations around saddle-point:}

We thus first focus on the half-density case.
One looks for a space-time perturbation around the saddle-point solutions of the form
\begin{equation}
  \rho_k(t)=\rho_k^* + N^{-\frac 12}\phi_k(t),
\qquad
  \hat\rho_k(t)=\hat\rho_k^* + N^{-\frac 12} \hat\phi_k(t),
\label{eq:deffluctphi}
\end{equation}
with $\phi_k(t)$ and $\hat\phi_k(t)$ of order $N^0$. 
The prefactor $N^{-\frac 12}$ is chosen so that when substituting~\eqref{eq:deffluctphi} into the action for $\brho$, $\hbrho$ the temporal contribution to the action is $-\int_0^Tdt\,\hat{\boldsymbol \phi}\cdot\partial_t \boldsymbol \phi$, whose absence of prefactor facilitates further analysis.
%
%
Expanding in powers of $N$, the total Hamiltonian~\eqref{eq:spep_H} decomposes as
\begin{equation}
H_L(\lambda;\brho,\hbrho)
=
\underbrace{
H_\lambda^\text{saddle}
}_{\textnormal{order $N^{0}$}}
+ 
\underbrace{
  H_\lambda^\text{fluct}(\boldsymbol \phi, \hat {\boldsymbol \phi})
}_{\textnormal{order $N^{-1}$}}
\: + \ {\text{higher order terms}},
\end{equation}
with 
\begin{equation}
H_\lambda^\text{saddle}
=H_L(\lambda;\boldsymbol \rho^*, \hat {\boldsymbol \rho}^*)
=(L+1)\sinh^2\frac{\lambda}{2(L+1)}
\end{equation}
yielding the dominant contribution $\psi_L(\lambda)$ to the full CGF $\psi_{N,L}$. Meanwhile, $H_\lambda^\text{fluct}(\boldsymbol \phi, \hat {\boldsymbol \phi})$ has a quadratic form
\begin{equation}
  H_\lambda^\text{fluct}(\boldsymbol \phi, \hat {\boldsymbol \phi})
  = 
N^{-1}
\big(\begin{smallmatrix}
  \boldsymbol \phi\\ \hat {\boldsymbol \phi}
\end{smallmatrix}\big)^T 
\mathcal A \:
\big(\begin{smallmatrix}
  \boldsymbol \phi\\ \hat {\boldsymbol \phi}
\end{smallmatrix}\big),
\label{eq:defmatAHfluct}
\end{equation}
where $\mathcal A$ is a symmetric $2L\times 2L$ matrix defined by block structure
\begin{equation}
  \mathcal A = 
  \begin{pmatrix}
    \mathcal A_{11} &\mathcal A_{12} \\ \mathcal A_{21}&\mathcal A_{22}
  \end{pmatrix}
\label{eq:decompositionA2x2blocks}
\end{equation}
with the $\mathcal A_{ij}$'s symmetric $L\times L$ matrices
\begin{align}
  \mathcal A_{11}
 & = 2 \big(\tfrac12 \Delta + \mathbf 1\big) (1-\cosh\frac{\lambda}{L+1})
\\
  \mathcal A_{12}=\mathcal A_{21}
  &=\tfrac 12 \Delta\cosh\frac{\lambda}{L+1}
\\
  \mathcal A_{22}=-\tfrac 12 \mathcal A_{12}
  &=-\tfrac 14 \Delta\cosh\frac{\lambda}{L+1}.
\end{align}
Here the $L\times L$ matrix $\Delta$ is the discrete Laplacian (with open boundaries)
\begin{small}
  \begin{equation}
    \Delta
    =
    \begin{pmatrix}
      -2&1 &0& &\ldots&0 \\
      1&-2&1&0&\ldots&0 \\
      0&\ddots&\ddots&\ddots&\ddots&0 \\
      0&\ddots&\ddots&\ddots&\ddots&0 \\
      0&\ldots&0&1&-2&1 \\
      0&\ldots&&0&1&-2 \\
    \end{pmatrix}.
  \end{equation}
\end{small}
The eigenvalues of $\Delta$ are
\begin{equation}
  \Delta_p = - 4 \cos^2\frac{p\pi}{2(L+1)}\;,
\qquad
1\leq p\leq L,
\end{equation}
with corresponding orthonormal eigenvectors
\begin{equation}
  \mathbf V_p= \sqrt{\tfrac{2}{L+1}}\Big(\sin\frac{(L+1-p) k \pi}{L+1}\Big)_{1\leq k\leq L}.
\label{eq:defVpeigenvectorsDelta}
\end{equation}

\subsection{Corrections due to ``quantum fluctuations'': mapping to independent bosons}

At the quadratic order, one has
\begin{equation}
e^{NT\psi_{N,L}(\lambda)} \simeq 
e^{N T H_\lambda^\text{saddle}}
\int\mathcal D\boldsymbol\phi \mathcal D \hat{\boldsymbol\phi}
\
e^{
-\int_0^T dt
\big[
\hat{\boldsymbol\phi}\cdot\partial_t\boldsymbol\phi
-N H_\lambda^\text{fluct}(\boldsymbol \phi, \hat {\boldsymbol \phi})
\big]\;.
}
\label{eq:PIQF}
\end{equation}
To compute the path integral and evaluate the so-called ``quantum fluctuations'', one can regard \eqref{eq:PIQF} as a coherent-state path integral of a bosonic harmonic oscillator, whose ground state becomes dominant in the large $T$ limit. This leads to
\begin{equation}
  \psi_{N,L}(\lambda) = \psi_L(\lambda) - N^{-1} \min \operatorname{Sp} \mathbf H_\lambda + o(N^{-1}).
 \label{eq:relationpsiLHoplambda}
\end{equation} 
The operator $\mathbf H_\lambda$ is such that its coherent-state path integral is given by \eqref{eq:defmatAHfluct}; thus, it can be written in the form
\begin{equation}
-\mathbf H_\lambda = 
 \boldsymbol a^T \mathcal A_{11} \boldsymbol a+2{\boldsymbol a^\dag}^T \mathcal A_{12} \boldsymbol a+{\boldsymbol a^\dag}^T \mathcal A_{22} {\boldsymbol a^\dag},
%
%
\label{eq:defHopAij}
\end{equation}
where the operators $a_1,\ldots,a_L$ are bosonic annihilation operators and $a^\dagger_1,\ldots,a^\dagger_L$ are their creation counterparts, with $[a_i,a_j^\dagger]=\delta_{ij}$.
We remark that choosing a scaling other than $N^{-\frac 12}$ for the fluctuations in~\eqref{eq:deffluctphi} would leave the result~(\ref{eq:relationpsiLHoplambda}-\ref{eq:defHopAij}) \emph{unchanged} (the only important aspect being that the exponent is negative, allowing for a perturbation expansion).

The eigenvectors~\eqref{eq:defVpeigenvectorsDelta} define an orthonormal matrix $O$ which renders the modes independent.
Using
\begin{equation}
  O \Delta O^T = \operatorname{Diag}(\Delta_1,\ldots,\Delta_L)\equiv \widetilde\Delta,
\end{equation}
one obtains
\begin{align}
-\mathbf H_\lambda 
&= 
 \boldsymbol a^TO^TO \mathcal A_{11}O^TO \boldsymbol a+\ldots
%
\\
&= 
 \boldsymbol b^T \widetilde {\mathcal A}_{11} \boldsymbol b+2{\boldsymbol b^\dag}^T \widetilde {\mathcal A}_{12} \boldsymbol b+{\boldsymbol b^\dag}^T \widetilde {\mathcal A}_{22} {\boldsymbol b^\dag},
\end{align}
where $\boldsymbol b = O \boldsymbol a$ are new bosonic annihilation operators, and $\widetilde {\mathcal A}$ consists of four blocks as in~\eqref{eq:decompositionA2x2blocks}, with each block $\widetilde {\mathcal A}_{ij}$ being a symmetric $L\times L$ matrix given by
\begin{align}
\widetilde {\mathcal A}_{11}
 & = 2 \big(\tfrac12 \widetilde\Delta + \mathbf 1\big) (1-\cosh\frac{\lambda}{L+1}),
\\
\widetilde {\mathcal A}_{12}=\widetilde {\mathcal A}_{21}
  &=\tfrac 12 \widetilde\Delta\cosh\frac{\lambda}{L+1},
\\
\widetilde {\mathcal A}_{22}=-\tfrac 12 \widetilde {\mathcal A}_{12}
  &=-\tfrac 14 \widetilde\Delta\cosh\frac{\lambda}{L+1}.
\end{align}
Because each $\widetilde {\mathcal A}_{ij}$ is diagonal, the operator $\mathbf H_\lambda$ can be written as a sum of independent single-boson operators
\begin{align}
\mathbf H_\lambda &= \sum_{p=1}^L \mathbf H_\lambda^{(p)},
\\
 \mathbf H_\lambda^{(p)}
&=
\underbrace{
- 2 \big(\tfrac12 \Delta_p + 1\big) (1-\cosh\frac{\lambda}{L+1})}
_
{\equiv X_p}
b_p^2
\underbrace{
-
\Delta_p\cosh\frac{\lambda}{L+1}}
_
{\equiv 2 Z_p}
 b^\dag_pb_p
+
\underbrace{
\tfrac 14 \Delta_p\cosh\frac{\lambda}{L+1}
}
_
{\equiv Y_p}
 \big(b^\dag_p\big)^2.
\end{align}
Besides (\emph{e.g.},~through a generalized Bogoliubov transform), one finds that the ground state of every $\mathbf H_\lambda^{(p)}=X_pb_p^2+2Z_pb^\dag_pb_p+Y_p\big(b^\dag_p\big)^2$
(seen as a harmonic oscillator) is given by:
\begin{equation}
  \min \operatorname{Sp} \mathbf H_\lambda^{(p)}
=
\sqrt{Z_p^2-X_pY_p}-Z_p.
\end{equation}

We finally obtain the result for $\bar\rho_a=\bar\rho_b=\frac 12$.
Denoting $\boldsymbol c_\lambda=\cosh \frac{\lambda}{L+1}$, the correction $\psi_L^1(\lambda)$ in~\eqref{eq:spep_cgf_fin_N_series} due to space-time fluctuations reads
\begin{align}
\psi_L^1(\lambda)
&=
 -\min  \operatorname{Sp} \mathbf H_\lambda
\\
&= 
 -\sum_{p=1}^L \min  \operatorname{Sp} \mathbf H_\lambda^{(p)}
\\
&=
-\sum_{p=1}^L
\bigg\{
 \cos\frac{p\pi}{2(L+1)}
\sqrt{2\boldsymbol c_\lambda\big(\boldsymbol c_\lambda+\cos\frac{p\pi}{L+1}\big)}
 - 2 \boldsymbol c_\lambda  \cos^2\frac{p\pi}{2(L+1)}
\bigg\}
\label{eq:respsiLhalfdensities_app}
\end{align}
For generic reservoir densities $\bar \rho_a$ and $\bar \rho_b$, 
one can use the mapping~\eqref{eq:mmappingpsiLfiniteN} to find that
the correction term still takes the form of~\eqref{eq:respsiLhalfdensities_app}, but now with
\begin{equation}
  \boldsymbol c_\lambda= \cosh
 \frac{2\operatorname{arcsinh}\sqrt{\omega}}{L+1}.
\end{equation}

Averaging the $p$-th and $(L+1-p)$-th terms, the sum~\eqref{eq:respsiLhalfdensities_app} can be symmetrized as
\begin{align}
\psi_L^1(\lambda)
&=
\sum_{p=1}^{L'-1}
\underbrace{
\bigg\{
\boldsymbol c_\lambda
-
\cos\frac{p\pi}{2L'}
\sqrt{\frac 12 \boldsymbol c_\lambda\big(\boldsymbol c_\lambda+\cos\frac{p\pi}{L'}\big)}
-
\sin\frac{p\pi}{2L'}
\sqrt{\frac 12 \boldsymbol c_\lambda\big(\boldsymbol c_\lambda-\cos\frac{p\pi}{L'}\big)}
\bigg\}
}_{\equiv \Sigma_{L'}(p)},
\label{eq:respsiLhalfdensities_sym_final}
\end{align}
where we used a notation $L'=L+1$.
This is our final result for the leading finite-$N$ correction to the CGF $\psi_L(\lambda)$. 

One checks that this expression is an analytic function of $\lambda$ at all system size $L$, indicating that the optimal profiles $\brho^*,\hbrho^*$ are stable with respect to any small perturbations in space and time. This is consistent with the hypothesis of additivity that we assumed to derive $\brho^*,\hbrho^*$, but there is still the possibility of discontinuous transitions (which, if they exist, can also be ruled out provided that they have a spinodal).

Taking the large $L$ limit is not straightforward because (\emph{i}) the summand $\Sigma_{L'}(p)$ in~\eqref{eq:respsiLhalfdensities_sym_final} exhibits different scaling with $L$ depending on the value of $p$, and (\emph{ii}) the range of $p$ itself depends on $L$. In particular, the sum cannot be approximated by a Riemann integral because the summand $\Sigma_{L'}(p)$, seen as function of a continuous variable $p \in (0,L')$, is not an analytic function.
In fact, \eqref{eq:respsiLhalfdensities_sym_final} remains a discrete sum even in the large-$L$ limit, as we now explain.

We first note that for $L'$ even, $\Sigma_{L'}(L'/2)=0$. For any $L'$, thanks to the symmetry $\Sigma_{L'}(p)=\Sigma_{L'}(L'-p)$, one can thus restrict the sum as follows:
\begin{align}
\psi_L^1(\lambda)
&=
2\sum_{p=1}^{\lfloor L'/2 \rfloor}
\Sigma_{L'}(p)
\end{align}
At fixed $\lambda$, the leading-order term in $L'\to\infty$ gives (provided $1\leq p\leq\lfloor L'/2 \rfloor$)
\begin{equation}
  \Sigma_{L'}(p) = 
  \frac{1}{L'^2}
  \frac 14
  \Big\{
    (p\pi)^2-\mu(\lambda)-p\pi\sqrt{(p\pi)^2+2\mu(\lambda) }
  \Big\}
\ + \
  O(L^{-3}),
\end{equation}
with
$\mu(\lambda)=\operatorname{arcsinh}^2\sqrt{\omega}$.
Then, using the Euler-Maclaurin summation formula to control the rest (\emph{i.e.},~the terms with $p>L'/2$), one finds
\begin{align}
  \psi_L^1(\lambda) 
  &= \frac{1}{2L^2}
  \sum_{p=1}^{\infty}
  \Big\{
    (p\pi)^2-\mu(\lambda)-p\pi\sqrt{(p\pi)^2+2\mu(\lambda) }
  \Big\}
  \ + \
  O(L^{-3})
\label{eq:respsi1L_largeL}
\\
&=
  \frac 1{8L^2} \mathcal F\big(\!-\!\mu(\lambda)\big) \ + \
  O(L^{-3}),
\end{align}
where we recognize the universal scaling function
\begin{align}
  \mathcal F(u) 
  &= 4 \sum_{p=1}^{\infty}
  \Big\{
   (p\pi)^2+u-p\pi\sqrt{(p\pi)^2-2u }
   \Big\}
\end{align}
appearing in MFT and Bethe-Ansatz studies of current fluctuations~\cite{appert-rolland_universal_2008,prolhac_cumulants_2009}.
The large-$L$ limit (at fixed $\lambda$) thus yields the same correction~\eqref{eq:respsi1L_largeL} as does the MFT approach~\cite{Imparato2009} for the SSEP.
The finite-$L$ result~\eqref{eq:respsiLhalfdensities_sym_final} however allows one to study large deviations regimes with $\lambda$ increasing as a function of $L$, which are not described by~\eqref{eq:respsi1L_largeL}.

Another illustration is obtained by a direct expansion of the full result~\eqref{eq:respsiLhalfdensities_app} in powers of $\lambda$ at finite $L'$.
A direct summation on $p$ then yields (focusing without loss of generality on the case $\bar\rho_a=\bar\rho_b=\tfrac 12$)
\begin{align}
  \psi^1_L(\lambda,\bar\rho_a=\bar\rho_b=\tfrac 12)
=
\
&\left[
\frac{1}{3 {L'} ^4}-\frac{1}{2 {L'} ^3}+\frac{1}{6 {L'} ^2}
\right]
\frac{\lambda^4}{16}
\nonumber
\\
+
&\left[
\frac{1}{5 {L'} ^6}-\frac{1}{2 {L'} ^5}+\frac{1}{3 {L'} ^4}-\frac{1}{30 {L'} ^2}
\right]
\frac{\lambda^6}{96}
\nonumber
\\
+&\left[
\frac{25}{168 {L'} ^8}+\frac{9}{80 {L'} ^7}+\frac{3}{20 {L'} ^6}-\frac{11}{80 {L'}
   ^4}+\frac{1}{42 {L'} ^2}
\right]
\frac{\lambda^8}{1152}
+ O(\lambda^{10}).
\label{eq:expansionlambdafiniteLpsi1}
\end{align}
The dominant terms, of order $1/{L'}^2$, correspond as expected to the expansion in powers of $\lambda$ of the large-$L$ result~\eqref{eq:respsi1L_largeL}.
The other terms are the one provided at finite $L$ by the full expression~\eqref{eq:respsiLhalfdensities_sym_final}.
We note that if one scales $\lambda$ with $L$ as $\lambda\sim L^\zeta$ ($\zeta>0$), the expansion~\eqref{eq:expansionlambdafiniteLpsi1} remains well defined only for $\zeta<1$.
For $\zeta\geq 1$ there is thus a change of regime, as also occurs for the saddle-point contribution $\psi_L(\lambda)$ to the full CGF $\psi_{N,L}(\lambda)$ (see the corresponding discussion of the hydrodynamic behavior in Sec.~\ref{ssec:spep_hydro}).

\section{Symmetric zero-range process} \label{app:zrp}

As a simple example supporting our conjecture on the relation between unbounded $\sigma(\rho)$ and hydrodynamic behaviors of current large deviations, we examine the symmetric zero-range process (ZRP) on an open one-dimensional system. In this model, a particle hops between neighboring sites at a rate $u(n_k)$ that depends only on the number of particles $n_k$ at the site of departure. More precisely, the bulk dynamics are given by
\begin{equation}\label{eq:zrp_bulk_rates}
	(n_k,\,n_l) \xrightarrow{u(n_k)} (n_k - 1,\, n_l + 1) \qquad \text{for $l = k \pm 1$ with $k = 2,\ldots,L-1$},
\end{equation}
while the boundary dynamics are given by
\begin{alignat}{2}\label{eq:zrp_boundary_rates}
	n_1 &\xrightarrow{\alpha} n_1 + 1,	&\qquad n_1 &\xrightarrow{\gamma u(n_1)} n_1 - 1, \nonumber\\
	n_L &\xrightarrow{\delta} n_1 + 1,	& n_L &\xrightarrow{\beta u(n_L)} n_L - 1.
\end{alignat}
It can be shown~\cite{spohn_large_1991,Kipnis1999,Bertini2002} that the hydrodynamic behaviors of the model are characterized by boundary conditions
\begin{equation} \label{eq:zrp_zab}
	\bar z_a \equiv z(\bar\rho_a) = \frac{\alpha}{\gamma}, \quad \bar z_b \equiv = z(\bar\rho_b) = \frac{\delta}{\beta},
\end{equation}
and transport coefficients
\begin{equation} \label{eq:zrp_coeff}
	D(\rho) = z'(\rho), \quad \sigma(\rho) = 2z(\rho),
\end{equation}
where the fugacity $z(\rho)$ is an increasing function of the particle density $\rho$ (see \cite{Evans2005}, for example). We assume that $z(\rho)$ is not bounded from above; otherwise, a condensation transition occurs for sufficiently large current fluctuations~\cite{Harris2005,Harris2006,Hirschberg2015}, in which case we can no longer discuss the steady-state statistics of the currents. Given this assumption, $\sigma(\rho)$ is not bounded from above, so our conjecture predicts that the symmetric ZRP shows hydrodynamic behaviors for arbitrarily large current fluctuations. We check this prediction by comparing microscopic and hydrodynamic scaled CGFs for the time-averaged current, which are defined through
\begin{equation}
	e^{T\psi_L^\mathrm{ZRP}(\lambda)} \sim \langle e^{\lambda TJ} \rangle,
	\quad e^{(L+1)^2 T\psi^\mathrm{ZRP}(\lambda)} \sim \langle e^{\lambda (L+1)^2 TJ} \rangle,
\end{equation}
respectively. Note that $\langle \cdot \rangle$ denotes the average over all possible evolutions of the system during a time interval $t \in [0,\,T]$ in the former and $t \in [0,\,(L+1)^2T]$ in the latter. Since the exact microscopic expression was derived in \cite{Harris2005} as
\begin{equation} \label{eq:zrp_cgf_micro}
	\psi^\mathrm{ZRP}_L(\lambda) = \frac{(1 - e^{-\lambda})(e^{\lambda}\alpha\beta-\gamma\delta)}{\gamma+\beta+\beta\gamma(L-1)},
\end{equation}
here we present a derivation of the corresponding hydrodynamic expression only.

From \eqref{eq:H_mft} and \eqref{eq:zrp_coeff}, in the hydrodynamic limit the effective Hamiltonian of the symmetric ZRP is
\begin{equation} \label{eq:zrp_h}
	H^\mathrm{ZRP}[\rho,\hat\rho] = \int_0^1 \mathrm{d}x\, \left[ -z'(\rho)(\partial_x \rho)(\partial_x \hat\rho) + z(\rho) (\partial_x \hat\rho)^2 \right]
\end{equation}
with boundary conditions given by
\begin{equation} \label{eq:zrp_bcs}
	z(\rho(0)) = \bar z_a, \quad z(\rho(1)) = \bar z_b, \quad \hat\rho(0) = 0, \quad \hat\rho(1) = \lambda.
\end{equation}
Assuming an additivity principle, the optimal profiles satisfy
\begin{align} \label{eq:zrp_ap}
	\partial_t \rho = \frac{\delta H}{\delta \hat\rho} = \partial_x \left[ \partial_x z(\rho) - 2z(\rho)\partial_x \hat\rho \right] = 0, \nonumber\\
	\partial_t \hat\rho = -\frac{\delta H}{\delta \rho} = -z'(\rho) \left[ \partial_x^2 \hat\rho + (\partial_x \hat\rho)^2 \right] = 0,
\end{align}
which are solved by
\begin{align} \label{eq:zrp_profiles}
	z(\rho^*(x)) &= -\left[(e^\lambda - 1)x + 1\right] \, \left[(\bar z_a - \bar z_b e^{-\lambda})x - 1\right], \nonumber\\
	\hat\rho^*(x) &= \ln \left[ 1 + (e^\lambda - 1)x \right].
\end{align}
Thus the hydrodynamic scaled CGF is obtained as
\begin{equation} \label{eq:zrp_cgf_hydro}
	\psi^\mathrm{ZRP}(\lambda) = \frac{H[\rho^*,\hat\rho^*]}{L+1} = \frac{(e^\lambda - 1)\bar z_a - (1-e^{-\lambda})\bar z_b}{L+1}.
\end{equation}
From \eqref{eq:zrp_zab}, \eqref{eq:zrp_cgf_micro}, and \eqref{eq:zrp_cgf_hydro}, we obtain
\begin{equation}
	\lim_{L \to \infty} \frac{\psi^\mathrm{ZRP}_L(\lambda)}{\psi^\mathrm{ZRP}(\lambda)}
	= \lim_{L \to \infty} \frac{(e^\lambda - 1)\frac{\alpha}{\gamma} - (1-e^{-\lambda})\frac{\delta}{\beta}}{(e^\lambda - 1)\bar z_a - (1-e^{-\lambda})\bar z_b}
	= 1,
\end{equation}
which is true for any scaling of $\lambda$ with $L$. This confirms our prediction that the symmetric ZRP shows hydrodynamic behaviors for arbitrarily large current fluctuations.

We check whether the rationale behind our conjecture is also at work here. Following the procedure used for obtaining \eqref{eq:J_profiles_hydro} and \eqref{eq:J_profiles_hydro_approx}, we obtain
\begin{align}
	J &= \frac{1}{L+1} \int_0^1 \mathrm{d}x\, \left[ -z'(\rho^*)\partial_x \rho^* + 2 z(\rho^*) \partial_x \hat\rho^* \right]
	= \frac{\bar z_a - \bar z_b}{L+1} + \frac{2}{L+1} \int_0^1 \mathrm{d}x\, \left[z(\rho^*) \partial_x \hat\rho^* \right] \nonumber\\
	&\simeq \frac{2}{L+1} \int_0^1 \mathrm{d}x\, \left[z(\rho^*) \partial_x \hat\rho^* \right].
\end{align}
Thus $J$ beyond the naive hydrodynamic regime is dominated by $z(\rho^*)$ and $\partial_x \hrho^*$. Due to \eqref{eq:zrp_profiles}, these quantities satisfy
\begin{align}
	z(\rho^*) &\simeq \begin{cases}
		e^\lambda x(1 - \bar z_a x) &\text{ if $\lambda > 0$ and $\lambda \gg 1$,} \\
		e^{-\lambda} \bar z_b x (1-x) &\text{ if $\lambda < 0$ and $|\lambda| \gg 1$,}
	\end{cases} \nonumber\\
	\partial_x \hat\rho^* &= \frac{e^\lambda - 1}{1 + (e^\lambda - 1)x} \simeq \begin{cases}
		\frac{1}{x} &\text{ if $\lambda > 0$ and $\lambda \gg 1$,} \\
		\frac{1}{x-1} &\text{ if $\lambda < 0$ and $|\lambda| \gg 1$} \\
	\end{cases}
\end{align}
for $0 < x < 1$. Hence, a large $J$ is supported by a large $z(\rho^*)$, while $\partial_x \hat\rho^* = O(N^0)$ throughout the bulk region. We note that $\partial_x \hat\rho^*$ becomes arbitrarily large close to the boundaries, attaining the order of $L$ (corresponding to the threshold for non-hydrodynamic behaviors found in the SPEP) for $x \sim 1/L$ (for $\lambda > 0$) or $1- x \sim 1/L$ (for $\lambda < 0$). But one can easily see that $J$ has negligible contributions from these boundary regions compared to the bulk in the $L \to \infty$ limit. Therefore, the symmetric ZRP confirms our proposed scenario of how $J$ stays hydrodynamic for unbounded $\sigma(\rho)$.

\bibliography{current_LDF_large-N}
\end{document}